\documentclass[a4paper,oneside,11pt]{article}
\usepackage{epsfig,multicol}
\usepackage{graphicx}
\usepackage{amsmath,amssymb}
\usepackage{mathrsfs}
\usepackage{graphics,xcolor}
\usepackage{caption}
\usepackage{subcaption}

\numberwithin{equation}{section}

\begin{document}
\setcounter{page}{1}

\begin{center}
{\Large \bf Neutron-proton scattering and singular potentials}
\end{center}

\vskip 19.6ex

\begin{center}
Mahmut Elbistan$^1$\footnote{\tt elbistan@impcas.ac.cn}\,,\ \  
Pengming Zhang$^1$\footnote{\tt zhpm@impcas.ac.cn} 
\ \ and\ \ 
J\'anos Balog$^{1,2}$\footnote{\tt balog.janos@wigner.mta.hu}
\\
\vskip 3ex
$^1$ {\it Institute of Modern Physics, 
Chinese Academy of Sciences,\\ Lanzhou 730000, China}\\
\vskip 1ex
$^2${\it MTA Lend\"{u}let Holographic QFT Group, Wigner Research Centre} \\
{\it H-1525 Budapest 114, P.O.B. 49, Hungary}\\

\end{center}


\vskip 14.5ex

\noindent
We consider a Bargmann-type rational parametrization of the nucleon scattering
phase shifts. Applying Marchenko's method of quantum inverse scattering
we show that the scattering data suggest a singular repulsive core of the
potential of the form $2/r^2$ and $6/r^2$ in natural units, for the 
${}^3S_1$ and ${}^1S_0$ channels respectively.
The simplest solution in the ${}^3S_1$ channel
contains three parameters only but reproduces all features of
the potential and bound state wave function within one percent error.
We also consider the ${}^3S_1$-${}^3D_1$ coupled channel problem with the
coupled channel Marchenko inversion method.

\newpage


\newcommand{\beq}{\begin{eqnarray}}
\newcommand{\eeq}{\end{eqnarray}}
\newcommand{\ee}{\end{equation}}



\section{Introduction and motivation}

\begin{figure}
\begin{flushleft}
\leavevmode
\centerline{\includegraphics[width=8cm]{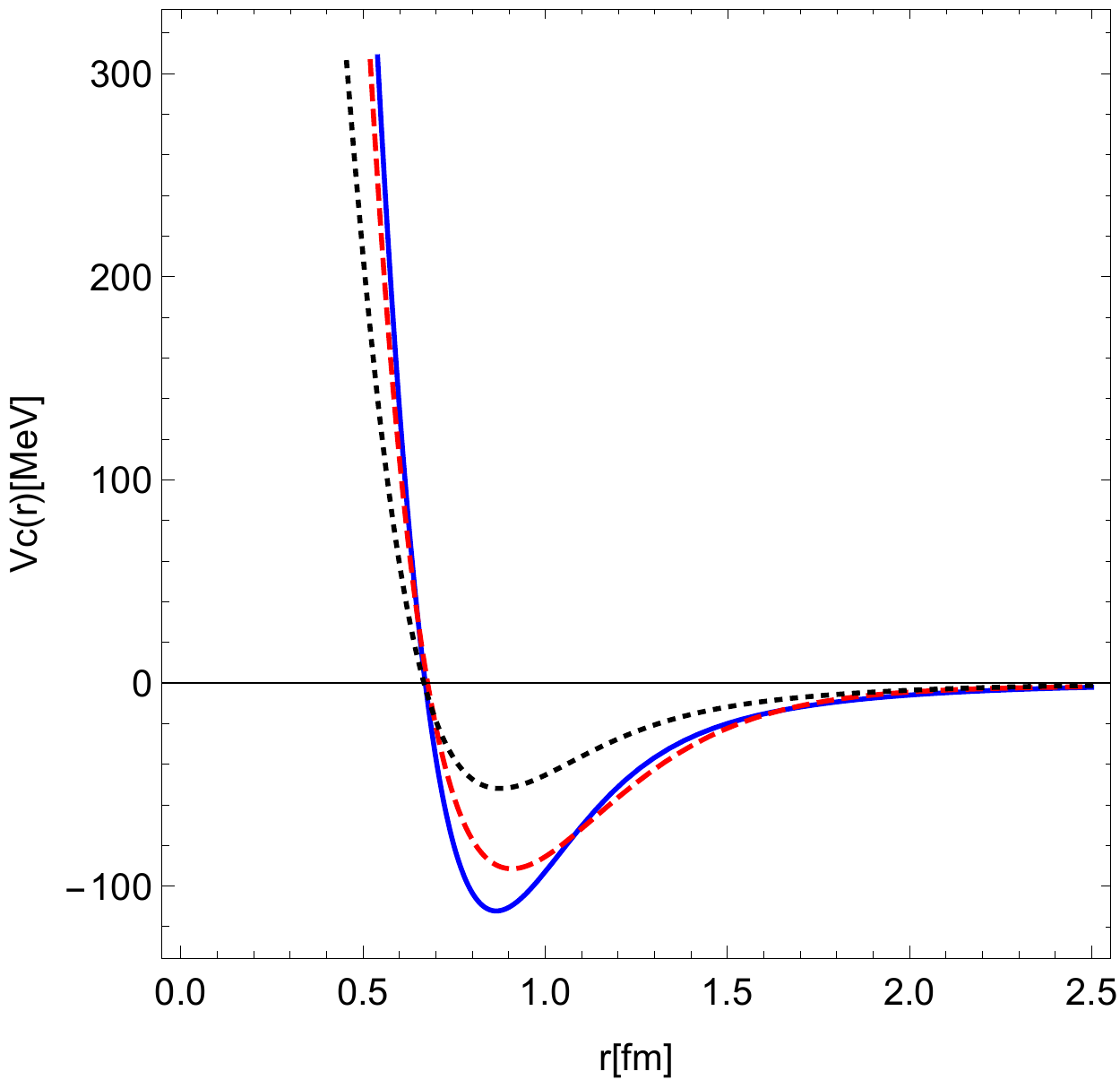} }
\end{flushleft}
\caption{{\footnotesize
The phenomenological nucleon potential in the ${}^1S_0$ channel.
The solid (blue) line is the AV18 fit \cite{NuPo3},
the dashed (red) line is the Reid93 fit \cite{NuPo2} and the
dotted line (black) is the CDBonn potential \cite{NuPo1}.
}}
\label{pheno}
\end{figure}

The phenomenological nucleon potential, shown in Fig. \ref{pheno}, is a
compilation of many decades' work of nuclear physicists
\cite{NuPo2,NuPo3,NuPo1}. Although the recent consensus is that the theory
of nuclear interactions must be based on effective field theory (for a
review, see \cite{EFT}), the phenomenological potential remains an important
source of intuition and often the starting point of quantitative work.

As can be seen in Fig. \ref{pheno}, the phenomenological potential is not
unique, since it is constructed to reproduce low energy scattering only.
Nevertheless, its main qualitative features are well-established.
The force at medium to long range is attractive; this feature
has long well been understood in terms of pion and other heavier meson
exchange. For a long time the characteristic repulsive core at short
distances had no satisfactory theoretical explanation, but with the advance
of lattice QCD simulations it became possible to determine the potential
in fully dynamical lattice QCD \cite{Ishii:2006ec,AokiRev}. The lattice
results resemble the phenomenological potential, including its repulsive
core, obtained for the first time from a first principles calculation.
The short distance behaviour of the potential can also be studied
in perturbative QCD, thanks to its asymptotic freedom. The results of the
perturbative calculations \cite{PT1,PT2} show that at extremely short
distances the potential behaves as $1/r^2$ (up to log corrections characteristic
to perturbative QCD). Recent calculations in holographic QCD \cite{HoloQCD}
also give an inverse square potential at short distances.

In this paper we study the simple Bargmann-type rational parametrization
of the S-matrix and determine the corresponding potential with Marchenko's
method of quantum inverse scattering \cite{QIS1c,QIS2}. In this approach
(see Appendix \ref{appB}) the potential is given by the formula
\beq
q(r)=-2\frac{{\rm d}^2}{{\rm d}r^2}\,\log{\cal D}(r),
\eeq
where ${\cal D}(r)$ is the determinant of a matrix with entries analytic in $r$.
If for small $r$ the determinant is approximately constant,
\beq
{\cal D}(r)={\cal D}_o+{\rm O}(r),
\eeq
then the corresponding potential is regular at the origin. However, if at $r=0$
the matrix becomes singular,
\beq
{\cal D}(r)={\rm O}(r),
\eeq
then
\beq
q(r)\sim\frac{\nu(\nu+1)}{r^2}
\eeq
with $\nu=1$.
Thus in this approach the $2/r^2$ type singularity naturally appears. If
${\cal D}(r)$ vanishes with a higher power of $r$, the strength of the
singularity $\nu$ is larger. Therefore this algebraic method is suitable
to study the singular core of the potential.

The structure of the paper is as follows. In sect. 2 we study the Bargmann-type
representation of the nucleon S-matrix and show that the phase shifts suggest
a singular potential. In sect. 3 we work out the simplest description in this
class in detail. This representation reproduces the phase shifts with a
maximal error of $1.5^\circ$ and also the deuteron parameters turn out to be
correct within 1 percent error.  In sect. 4 we present our results for the
${}^3S_1$-${}^3D_1$ coupled channel problem. In sect. 5 we study the
${}^1S_0$ channel and discuss the role of the singularity degree.
Our conclusions are summarized in sect. 6
while the technical details of our computations are given in the appendices.

\section{Bargmann-type parametrization of the nucleon potential}
\label{Bargmann}

Bargmann-type parametrization \cite{Barg} of the nucleon potential was already
used \cite{H1,H2} a long time ago to study nuclear scattering and bound state
problems. It has continued to be used ever since to study
scattering in various channels, and even for coupled channel
problems \cite{BP,SB,SS,X,vonGeramb1,vonGeramb2,SparBay}.
In recent studies the methods of SUSY QM (supersymmetric quantum mechanics)
has been used, which is also suitable for studying higher angular momentum and
coupled channel problems. For a review of SUSY methods in nuclear problems,
see \cite{review}.

The nucleon scattering  phase shifts vanish at zero energy. Originally,
also the vanishing of the phase shifts $\delta(k)$ (modulo $\pi$) for
scattering at large momentum $k$ and a corresponding regular potential
was assumed. Later it was realized that this class naturally contains also
$1/r^2$ type singular (repulsive) potentials. In this paper we will focus
on the singularity type of the potential and apply the original Marchenko
inverse scattering method. The advantage of the Marchenko method is that
it gives compact analytic expressions for the potential and the wave function.

The idea of applying Bargmann-type parametrization for the scattering matrix
$S(k)={\rm e}^{2i\delta(k)}$ originates from the study of the effective range
function\footnote{This formula for the effective range function is
valid for the s-wave only. For the d-wave 
we have to use the modified form (\ref{Rd}). }
\beq 
R(k)=ik\frac{S(k)+1}{S(k)-1}=k\cot\delta(k).
\eeq
$R(k)$ is an even, real-analytic function of $k$ in a neighbourhood of
the origin in the $k$ plane. Analytic continuation for larger momentum
values may encounter poles or branch points. Note that in particular, with
the exception of the origin, $R(k)$ has a pole where the phase shift
vanishes (mod $\pi$).

For small momentum, the first two terms
\beq
R(k)=-\frac{1}{a}+\frac{rk^2}{2}+vk^4+{\rm O}(k^6)
\label{ERE}
\eeq
in the low energy expansion give a satisfactory description of the
scattering in terms of two parameters: the scattering length $a$ and the
effective range $r$. For larger momentum values one can take into account
more terms in this expansion, but a more efficient description, reproducing
the poles at larger $k$ values, is provided by a Pad\'e type approximation
\beq
R(k)=\frac{M(k^2)}{N(k^2)},
\eeq
where $M$ and $N$ are both real polynomials, of degree $d_M$ and $d_N$,
respectively. 

This rational function approximation of $R(k)$ is equivalent to a
rational parametrization of the scattering matrix of the form
\beq
S(k)=\prod_{j=1}^{\cal N}\sigma_{z_j}(k),\qquad
\sigma_z(k)=\frac{z-ik}{z+ik},
\label{ratS}
\eeq
where the parameters $z_j$ are the roots of the algebraic equation
\beq
M(-z^2)+z N(-z^2)=0.
\eeq
We see that the property $S(0)=1$ is built in, but the phase shift at
large energies depends on the parity of the number of roots,
\beq
S(\infty)=(-1)^{\cal N}.
\eeq
If for large momentum
\beq
S(k)=1+{\rm O}(1/k),
\eeq
then the effective range function diverges for large $k$ like O$(k^2)$.
If however for large momentum
\beq
S(k)=-1+{\rm O}(1/k),
\eeq
then $R(k)$ approaches a constant at large $k$. Thus $[d_M/d_N]$, the type
of the Pad\'e approximation, is related to the large energy behaviour of the
phase shift and, due to Levinson's theorem, to the short distance singularity
of the potential. (See Appendix~\ref{appA}.)

In Ref. \cite{BP} the effective range function for neutron-proton scattering
in the triplet (${}^3S_1$) channel was parametrized as
\beq
R_t(k)=-\frac{1}{a_t}+\frac{r_tk^2}{2}+\frac{v_tk^4}{1-k^2/k_o^2}
\eeq
with
\beq
a_t=5.4030\,{\rm fm},\qquad\quad r_t=1.7494\,{\rm fm},
\quad\quad v_t=0.163\,{\rm fm}^3.
\label{vBP}
\eeq
The scattering length $a_t$ and the effective range $r_t$ were obtained from
the measured low energy phase shifts and the position of the pole at
$k=k_o=2.1057\,{\rm fm}^{-1}$ was determined from the vanishing of the phase
shift (see Fig. \ref{phase}). Finally, the shape parameter $v_t$ was fitted
to the measured phase shifts in the range of laboratory neutron energies
up to $350\,{\rm MeV}$.

This parametrization is a $[2/1]$ type Pad\'e fit, which can also be written
in the form
\beq
R_t(k)=\frac{w_0+w_1k^2+w_2k^4}{1-k^2/k_o^2},
\eeq
where (using fm units)
\beq
w_0=-0.1851,\qquad w_1=0.9164,\qquad w_2=-0.0343.
\eeq
The corresponding S-matrix has four roots $z_j$ (${\cal N}=4$). For
large $k$, $R_t(k)={\rm O}(k^2)$ and $S(\infty)=1$. However, the coefficient
$w_2$ is extremely small. We take it as an indication that we get a better
description if
\beq
R_t(k)\to{\rm const.},\qquad\quad S(\infty)=-1.
\label{minus}
\eeq
In ref. \cite{X} a $[2/2]$ type Pad\'e fit with ${\cal N}=5$ corresponding to
$S(\infty)=-1$ was studied. In the next section we use a $[1/1]$
type Pad\'e approximation, which is the simplest possibility with
property (\ref{minus}).

\section{Simplest Bargmann representations}
\label{simplest}

\subsection{A simple $[1/1]$ type Pad\'e fit}
\label{simplest1}

Motivated by the above observations we make a simple $[1/1]$ type
Pad\'e fit to the scattering data. 

\begin{figure}
\begin{flushleft}
\leavevmode
\centerline{\includegraphics[width=7cm]{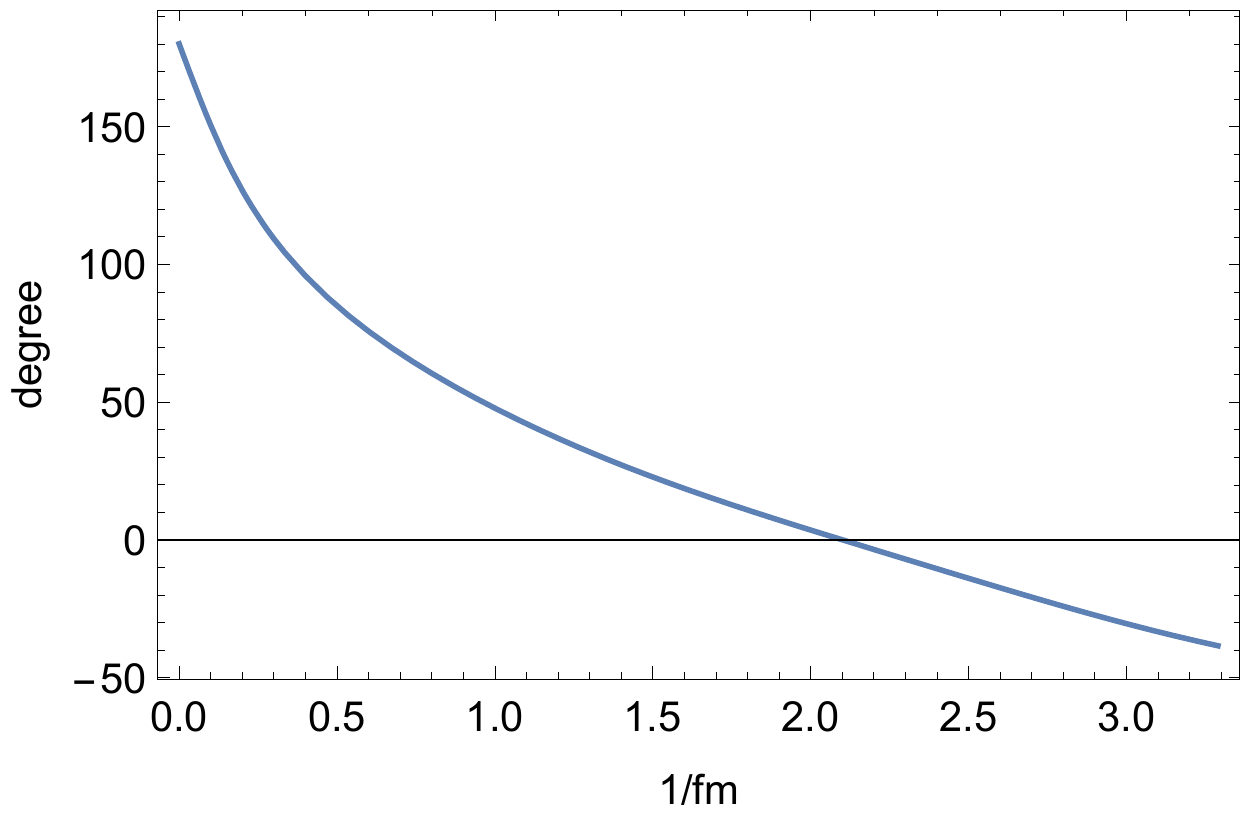} }
\end{flushleft}
\caption{{\footnotesize
    Neutron-proton scattering phase shifts in the ${}^3S_1$ channel as function
    of the centre of mass momentum \cite{DATA}. 
}}
\label{phase}
\end{figure}

We use the nucleon scattering phase shifts from the publicly available
GWDAC data base \cite{DATA} (SM16 solution).
The ${}^3S_1$ channel neutron-proton phase shifts are shown in Fig. \ref{phase}.
The phase shifts $\delta(k)$ are shown (in degrees) as function of the
centre of mass wave number $k$ corresponding to neutron laboratory kinetic
energies in the range $0\leq T\leq 900\,{\rm MeV}$. We will measure distances
(wave numbers) in fm (${\rm fm}^{-1}$) units, energies and potentials in MeV.

The phase shifts are decreasing with energy, starting from
$\delta(0)=180^\circ$, which corresponds to $S(0)=1$. The phase shift
vanishes at $T=366.28\,{\rm MeV}$, corresponding to
$k=k_o=2.1007\,{\rm fm}^{-1}$.
Here the phase shifts change sign and continue to decrease with energy.

Fitting to the low energy data in the range $0\leq T\leq 10\, {\rm MeV}$
we find the scattering length/effective range parameters\footnote{Our
results differ slightly from those of Ref. \cite{BP} because we use recent
experimental data.}
\beq
a_t=5.4028\,{\rm fm},\qquad\qquad r_t=1.7495\,{\rm fm}.
\label{standard}
\eeq
To make the $[1/1]$ type Pad\'e fit, we need no additional parameters beyond the
well-established set (\ref{standard}) and $k_o$. We find
\beq
R_t(k^2)=-\frac{1}{a_t}\frac{1-(a_t r_t/2+1/k_o^2)k^2}{1-k^2/k_o^2}=
-\frac{1}{a_t}\frac{1-4.9527\, k^2}{1-0.2266\,k^2}.
\label{Pade11}
\eeq
This gives for the next coefficient in the low energy expansion (\ref{ERE})
the value
\beq
v_t=\frac{r_t}{2k_o^2}=0.1982\, {\rm fm}^3,
\eeq
not very different from (\ref{vBP}). Using (\ref{Pade11}) gives a reasonable
fit to the phase shifts $\delta(k)$ up to laboratory energies
$350\,{\rm MeV}$, with errors not exceeding $1.4^\circ$.

The S-matrix corresponding to (\ref{Pade11}) contains three factors
\beq
S(k)=\sigma_a(k)\sigma_b(k)\sigma_c(k)
\label{Sabc}
\eeq
with $a=1.5813$, $b=0.23135$, $c=2.2327$.
We can interpret the S-matrix pole at $k=ib$ as corresponding to the deuteron
bound state since this $b$ value is very close to $b_o=0.2316$, obtained from
the deuteron binding energy
\beq
E_d=\frac{\hbar^2 b_o^2}{2m}=2.2245 \,{\rm MeV}.
\eeq
Here $m=469.459\, {\rm MeV}/c^2$ is the reduced mass of the $np$ system.

To characterize a system described by the rational S-matrix (\ref{ratS})
we will use the notation $({\cal N}_+,{\cal N}_-)[J]$, where ${\cal N}_\pm$
is the number of factors $\sigma_{z_j}$ with $z_j$ having positve (negative)
real parts and $J$ is the number of bound states.
Thus our simple fit is of type $(3,0)[1]$. Further, we can argue that the
would-be leading exponential ${\rm e}^{-2br}$ must be absent from the potential,
since this would give a much slower decay at infinity than expected (the
Yukawa tail of the potential dominated by pion exchenge). This will happen
if this S-matrix pole is an inner one (see Appendix \ref{appA}), in which
case no new bound state parameters need to be introduced. 

\begin{figure}
        \centering
        \begin{subfigure}[b]{0.32\textwidth}
                \centering
                \includegraphics[width=\textwidth]{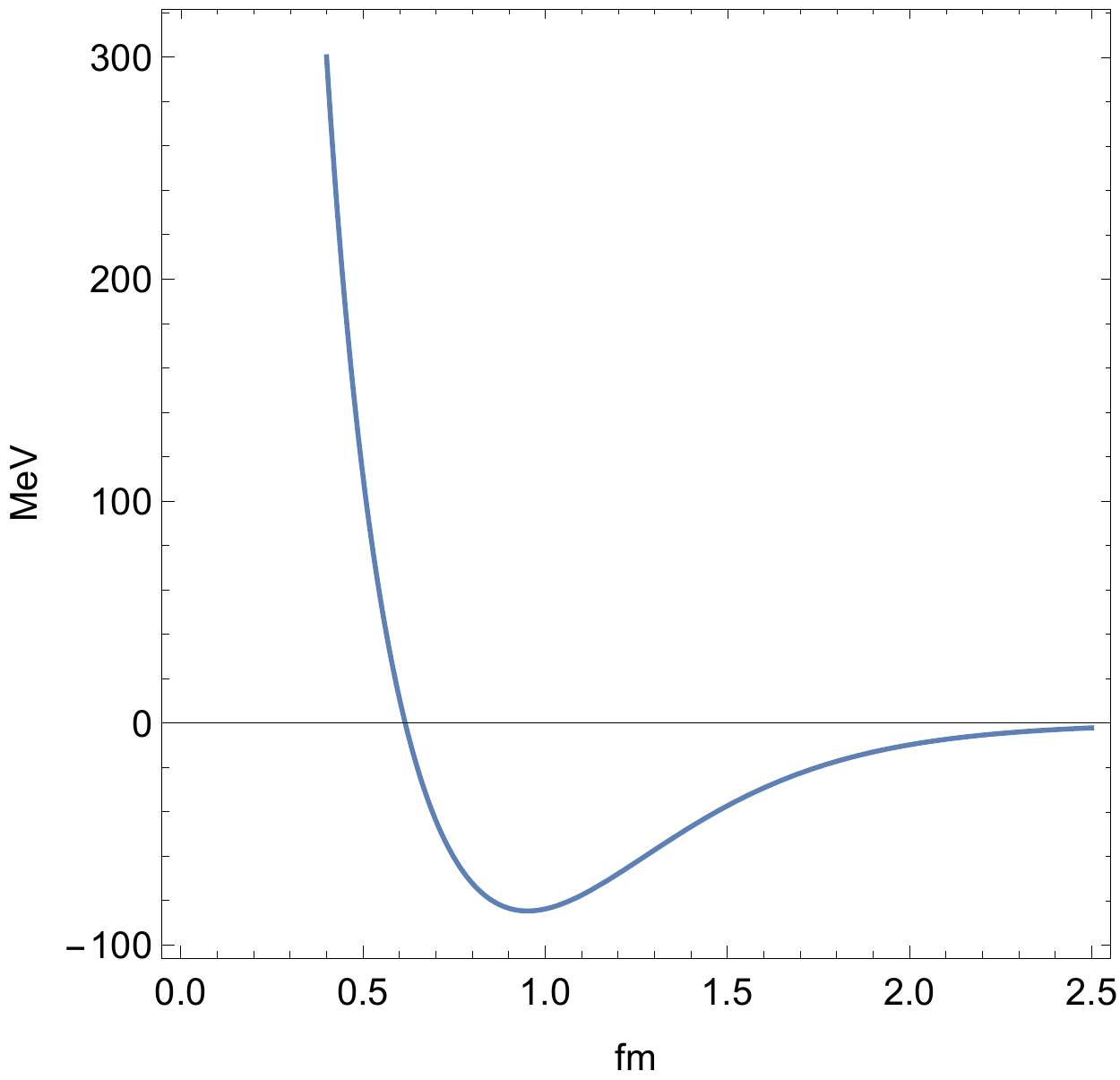}
                \caption{Our $(3,0)[1]$ type fit.}
                \label{3S111}
        \end{subfigure}%
        ~ 
        \begin{subfigure}[b]{0.3\textwidth}
                \centering
                \includegraphics[width=\textwidth]{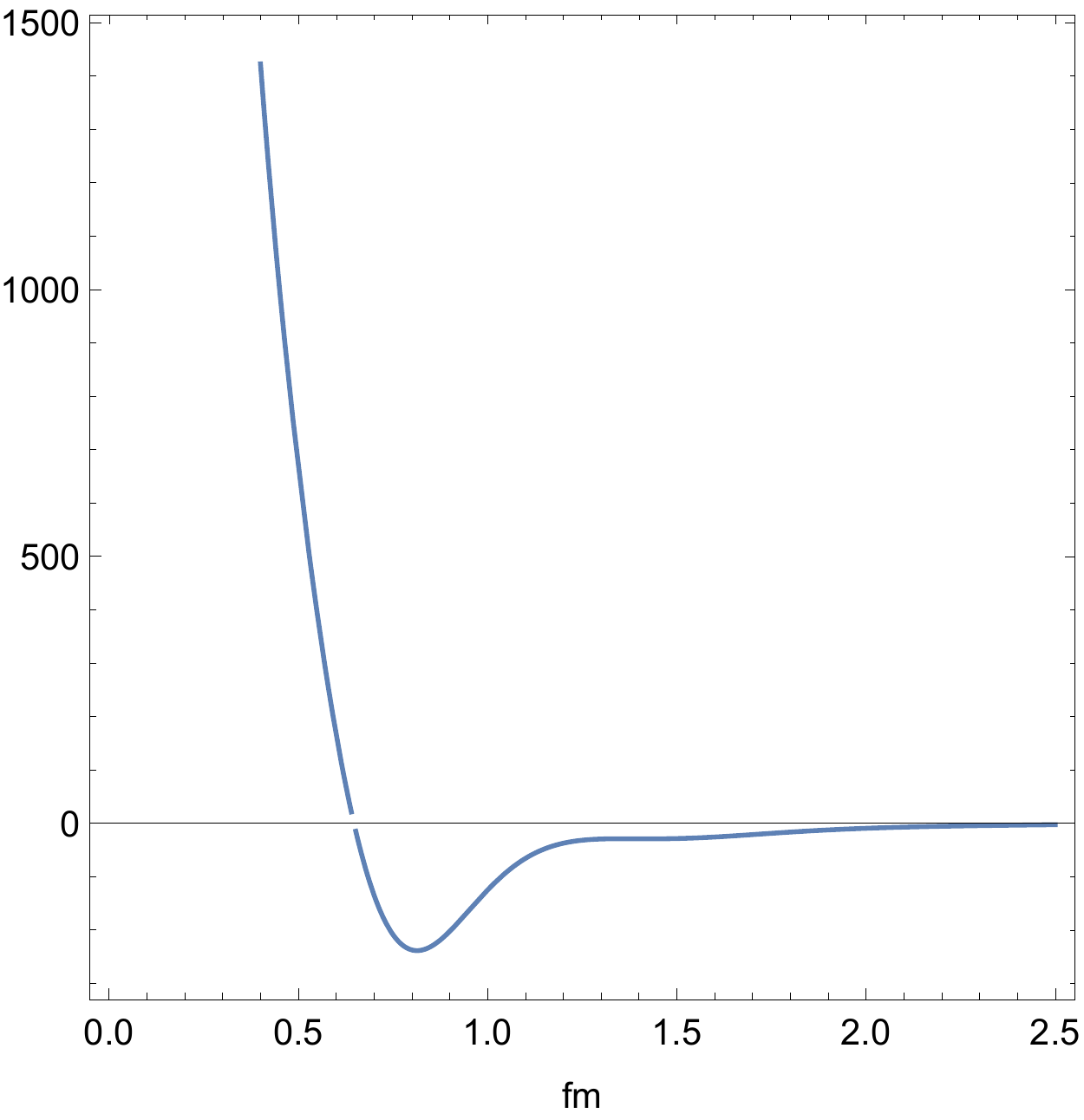}
                \caption{$(4,0)[1]$ type fit of \cite{BP}.}
                \label{3S1BP}
        \end{subfigure}
        ~ 
        \begin{subfigure}[b]{0.3\textwidth}
                \centering
                \includegraphics[width=\textwidth]{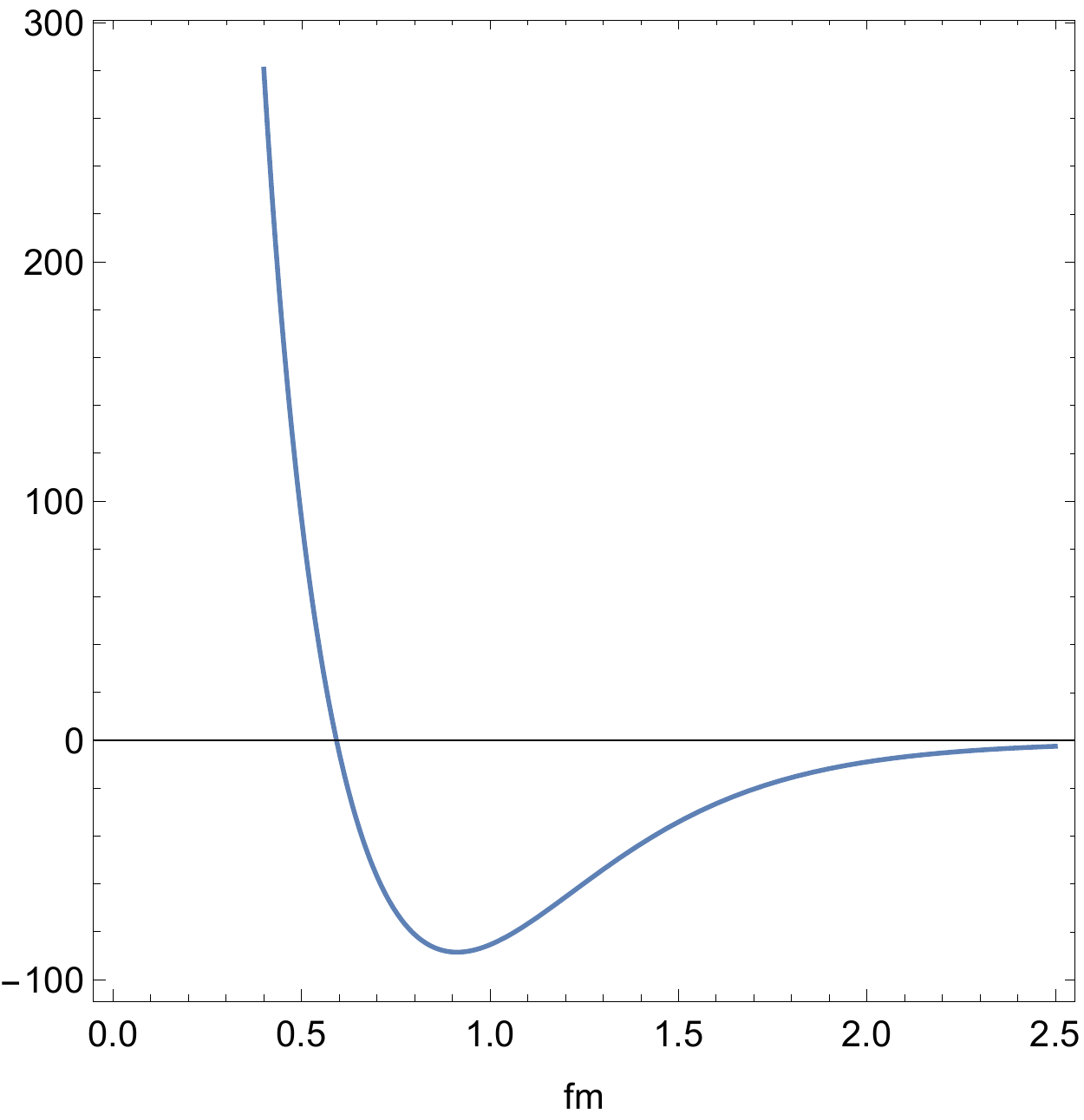}
                \caption{$(4,1)[1]$ type fit of \cite{X}.}
                \label{3S1X}
        \end{subfigure}
        \caption{Nucleon potentials from different Pad\'e fits. Note that
          the scale used in panel (b) is different from the other two
          due to the very different shape and range of the potential.}
\label{3S1fits}
\end{figure}

The solution of the inverse scattering problem corresponding to (\ref{Sabc})
is discussed in Appendix \ref{B2}. The solution is based on a $2\times2$ matrix
inversion and can be given analytically.

Let us define
\beq
Y_a(r)=\frac{(b-a)(c-a)}{(b+a)(c+a)}{\rm e}^{2ar}-1,\qquad
Y_c(r)=\frac{(a-c)(b-c)}{(a+c)(b+c)}{\rm e}^{2cr}-1
\eeq
and
\beq
\tilde {\cal D}(r)=Y_a(r)Y_c(r)-\frac{4ac}{(a+c)^2}.
\eeq
Then the potential (in fm units) is
\beq
q(r)=-2\frac{{\rm d}^2}{{\rm d}r^2}\ln\vert\tilde{\cal D}(r)\vert.
\eeq
It is shown in
Fig. \ref{3S111}. It closely resembles the phenomenological potential. It
has a single minimum $-84.7\,{\rm MeV}$ at $r=0.952\,{\rm fm}$ and is
singular for $r\to0$ behaving as
\beq
V(r)\sim \frac{\hbar^2}{2m}
\left(\frac{2}{r^2}-4.95+{\rm O}(r)\right)\,=
41.47\left(\frac{2}{r^2}-4.95+{\rm O}(r)\right)\,{\rm MeV}.
\eeq

We also calculated the ground state wave function.
\beq
\psi_o(r)=\sqrt{\tilde R_b}f_o(r),
\qquad
f_o(r)={\rm e}^{-br}\left\{1+\frac{4acB}{\tilde{\cal D}}\right\},
\eeq
where
\beq
\tilde R_b=2b\frac{(a+b)(c+b)}{(a-b)(c-b)}
\eeq
and
\beq
B=\frac{1}{a+b}\left\{\frac{Y_c}{2c}+\frac{1}{a+c}\right\}
+\frac{1}{c+b}\left\{\frac{Y_a}{2a}+\frac{1}{a+c}\right\}.
\eeq
In addition to the asymptotic constant $A_o$
characterising the large distance decay of the wave function,
\beq
\psi_o(r)\sim A_o{\rm e}^{-br},
\eeq
we have also calculated the root-mean-square \lq\lq matter'' radius of the
deuteron, defined by
\beq
(2r_m)^2=\int_0^\infty\,r^2\psi_o^2(r){\rm d}r.
\eeq
We find
\beq
A_o=0.8746\,{\rm fm}^{-1/2},\qquad\quad r_m=1.9441\,{\rm fm}.
\eeq
Again, we find very good agreement\footnote{Actually, the agreement is
even better with an older determination of these constants:\\
$A_o=0.8781(44)$,\ $r_m=1.9580(61)$ \cite{BP2}.}
with experimental data \cite{EXP},
\beq
A_o^{\rm exp}=0.8845(8)\,{\rm fm}^{-1/2},
\qquad\quad r_m^{\rm exp}=1.9676(10)\,{\rm fm}.
\eeq

\subsection{Higher Pad\'e approximations}
\label{simplest2}

For comparison, we have also constructed the potential and wave function
corresponding to the fits in Refs. \cite{BP} and \cite{X}.

The fit in Ref. \cite{BP} is of the form
\beq
R_t(k^2)=-\frac{1}{a_t}\frac{(1-4.9138\, k^2)(1-0.0377\, k^2)}{1-0.2255\,k^2},
\label{RtBP}
\eeq
reproducing the low energy expansion parameters (\ref{vBP}).

This $[2/1]$ type Pad\'e fit leads to the S-matrix
\beq
S(k)=\sigma_a(k)\sigma_b(k)\sigma_c(k)\sigma_d(k)
\label{Sabcd}
\eeq
with $a=1.2293$, $b=0.2315$, $c=2.5603+3.5248i$, $d=c^*$.
Identifying $b$ again with the deuteron pole and assuming the absence
of this exponent from the potential leads to an inverse scattering
problem of type $(4,0)[1]$. (See Appendix \ref{B3}.)

Although (\ref{RtBP}) gives better overall fit to the scattering
data (the errors in the phase shifts in the laboratory energy range
$0\leq T\leq 350\,{\rm MeV}$ are less than $0.5^\circ$), the resulting
potential, shown in Fig. \ref{3S1BP}, is very different from the
phenomenological one. It is much steeper, has a deep minimum 
$-238.8\,{\rm MeV}$ at $r=0.814$ and is more singular at the origin:
\beq
V(r)\sim 41.47\left(\frac{6}{r^2}+4.11+{\rm O}(r)\right)\,{\rm MeV}.
\eeq

We have also constructed the deuteron wave function and calculated the
parameters
\beq
A_o=0.8763\,{\rm fm}^{-1/2},\qquad\quad r_m=1.9442\,{\rm fm}.
\eeq

The $(4,1)[1]$ type S-matrix (see Appendix \ref{B3}) of Ref. \cite{X}
has parameters
\beq
\label{srootsX}
a=-0.45146,\quad b=0.23154,\quad c=0.43654,\quad
d=1.6818,\quad e=2.3106
\eeq
giving the $[2/2]$ type
\beq
R_t(k^2)=-\frac{1}{a_t}\frac{(1-4.8846\, k^2)(1+4.8594\, k^2)}
{(1-0.2104\,k^2)(1+4.9431\,k^2)}
\label{RtX}
\eeq
and low energy expansion parameters
\beq
a_t=5.4220\,{\rm fm},\qquad\quad r_t=1.7550\,{\rm fm},
\quad\quad v_t=0.0329\,{\rm fm}^3.
\eeq
The potential is shown in Fig. \ref{3S1X} and it is again very similar to
the phenomenological one with minimum $-88.6\,{\rm MeV}$ at $r=0.913$ and
short distance asymptotics
\beq
V(r)\sim 41.47\left(\frac{2}{r^2}-5.40+{\rm O}(r)\right)\,{\rm MeV}.
\eeq
The deuteron parameters are
\beq
A_o=0.8854\,{\rm fm}^{-1/2},\qquad\quad r_m=1.9568\,{\rm fm}.
\eeq

\section{Coupled channel nucleon-nucleon potential}
\label{coupled}

In section \ref{simplest1} we presented a Bargmann type S-matrix which has only three
poles (\ref{Sabc}). As can be seen from Fig. \ref{3S111}, this representation describes
the ${}^3S_1$ channel of the neutron-proton scattering quite well.  

However, the complete problem of triplet neutron-proton scattering includes
both the ${}^3S_1$ and ${}^3D_1$ channels which are coupled to each other.
In order to solve it completely,
we also need the data for the d-wave phase shifts and the mixing angle of the channels.
In addition, we have to employ the multichannel Marchenko method
\cite{vonGeramb1,vonGeramb2} for this inverse scattering (see appendix \ref{appC}).

Our 2-channel inversion procedure is based on the $2\times 2$ scattering matrix
$S(k)$ which can be decomposed as 
\beq
\label{multiS}
S(k)=O(k)\ {\rm{diag}}(e^{2i\delta_1(k)}, e^{2i\delta_2(k)})\ O^T(k), 
\label{para1}
\eeq
where $O(k)$ is an $SO(2)$ matrix,
\beq
O(k)=
\begin{pmatrix}
\cos\epsilon(k) & -\sin\epsilon(k) \\
\sin\epsilon(k) & \cos\epsilon(k) 
\end{pmatrix}.
\label{para2}
\eeq
$e^{2i\delta_1(k)}$ and $e^{2i\delta_2(k)}$ are the scattering matrices for the ${}^3S_1$ and ${}^3D_1$
channels, respectively.

For the ${}^3S_1$ channel scattering matrix we will use the results of section 3. First we take for $A(k)={\rm e}^{2i\delta_1(k)}$
(\ref{Sabc}) 
and later alternatively (\ref{Sabcd}) and the one parametrized by the roots (\ref{srootsX}). 

In order to determine $B(k)$ we downloaded the experimental ${}^3D_1$ phase shift data
from \cite{DATA}.
The simplest parametrization of the d-wave scattering matrix $B(k)={\rm e}^{2i\delta_2(k)}$
needs 5 poles because it is an $\ell=2$ partial wave and in the limit $k\to 0$
it has to satisfy $B(k)=1+O(k^5)$. Writing the effective range function as
\beq
R_d(k)=k^5\cot\delta_2(k)=ik^5\frac{B(k)+1}{B(k)-1}=c_o+c_1k^2+c_2k^4
\label{Rd}
\eeq
we conclude that there are only 3 independent parameters in the simplest solution for $R_d(k)$.
Fitting the three parameters in (\ref{Rd}), we obtain the following scattering phase:
\beq
\label{B}
B(k)=\sigma_a(k)\sigma_b(k)\sigma_c(k)\sigma_d(k)\sigma_e(k),
\eeq 
with poles at $a=-0.43936 - 0.47933i$, $b=a^*$,
$c=0.42490 - 0.52554i$, $d=c^*$, $e=4.5931$.
Like its s-wave counterpart, the d-wave scattering matrix (\ref{B}) has
the asymptotic value $B(\infty)=-1$. However, it has no bound states. 

\begin{figure}
        \centering
        \begin{subfigure}[b]{0.42\textwidth}
                \centering
                \includegraphics[width=\textwidth]{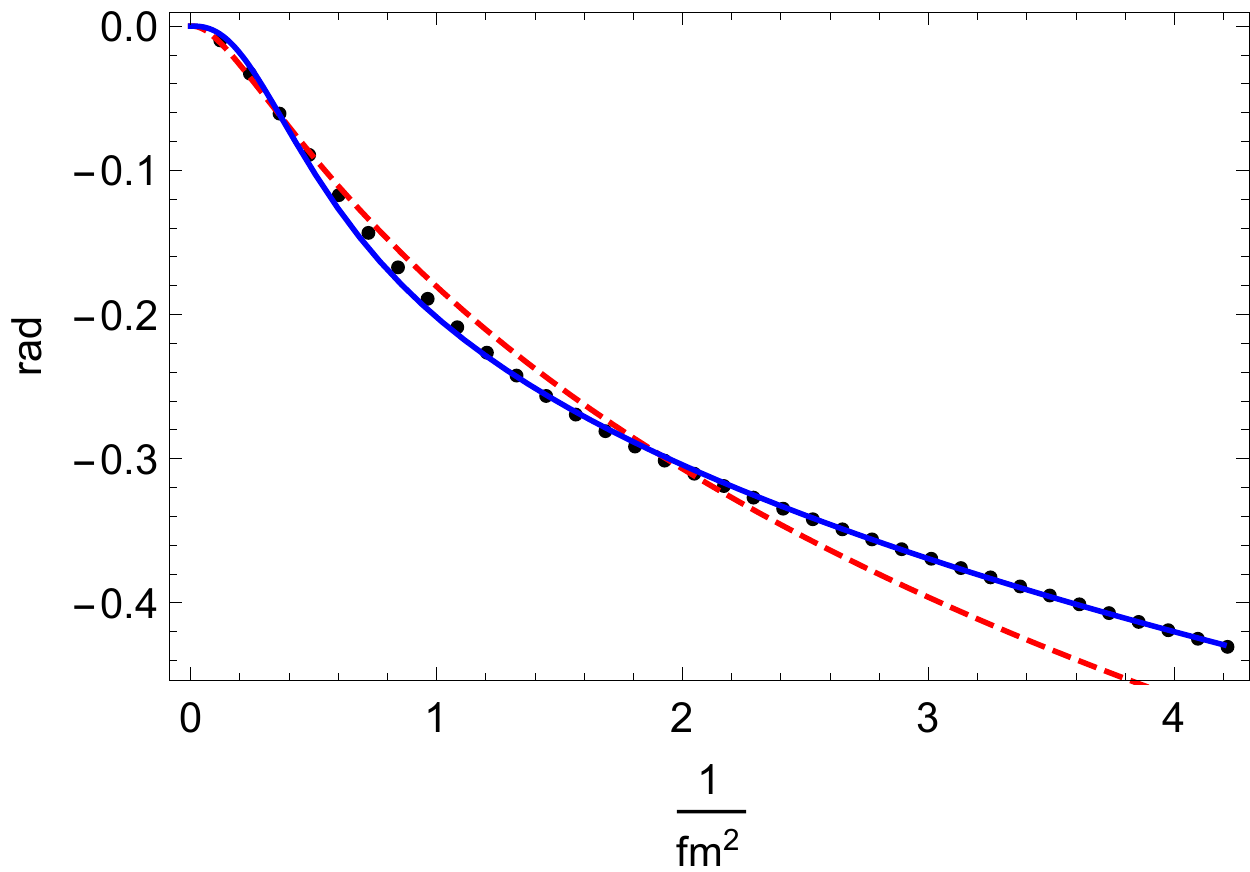}
                \caption{${}^3D_1$ channel phase shift $\delta_2$}
                \label{3D1}
        \end{subfigure} \quad \quad%
        ~ 
        \begin{subfigure}[b]{0.4\textwidth}
                \centering
                \includegraphics[width=\textwidth]{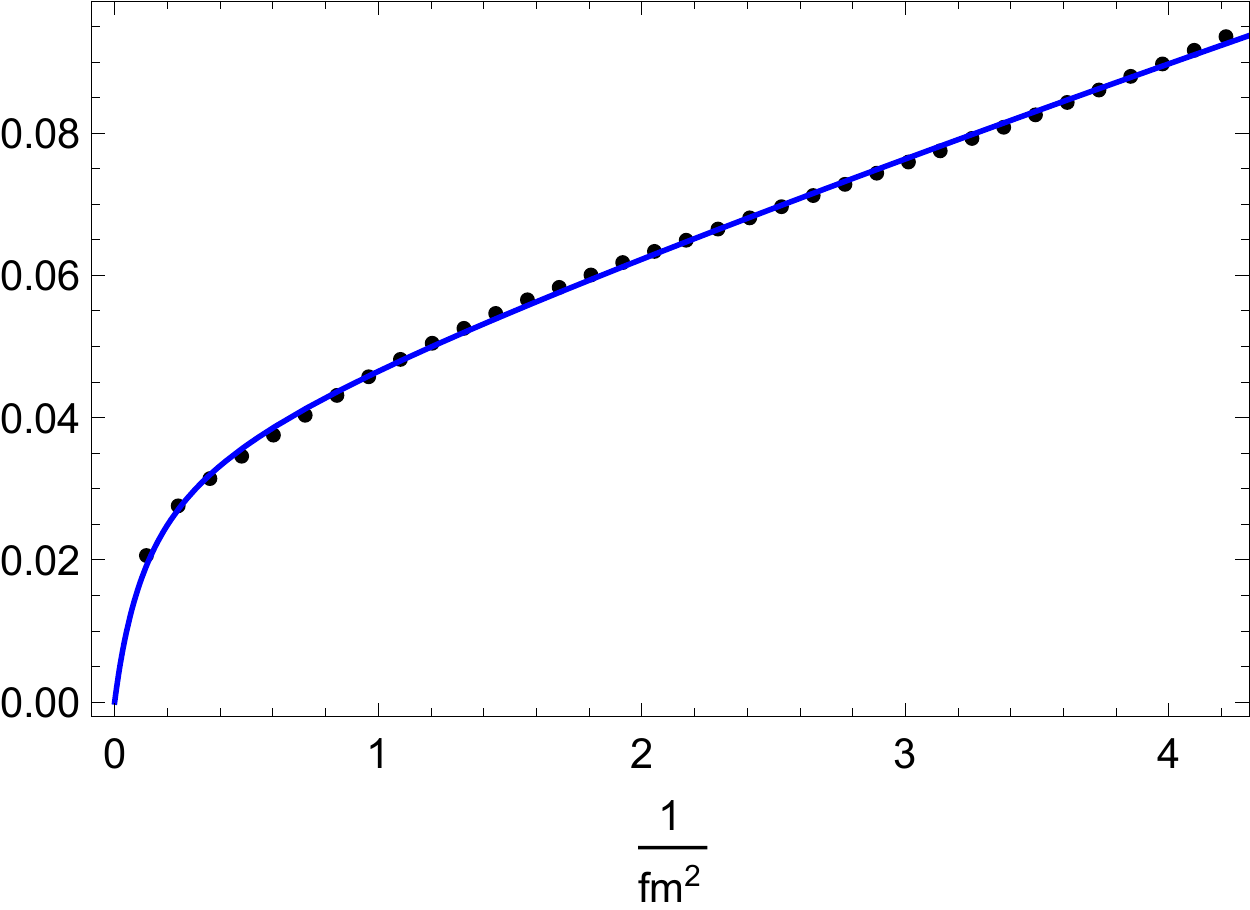}
                \caption{mixing angle $\epsilon$}
                \label{epsilon}
        \end{subfigure}
        ~ 
   \label{lalal}
\caption{$\delta_2$ and $\epsilon$ as
functions of the centre of mass momentum squared \cite{DATA}.
In both figures,  the solid (blue) line represents our fit and the dots represent the data. On the left, the dashed (red) line is the
fit used in \cite{X}. }
\end{figure}

In Fig. \ref{3D1}, we compare our d-wave phase shift fit  with the one in \cite{X}.
Although our poles are quite different, our fit is compatible with the one in \cite{X}. 

For the mixing angle $\epsilon (k)$, once again, we use the recent experimental data
from \cite{DATA}. We fit them to the functional form 
\beq
\label{mixingfit}
\epsilon(k)=\arctan\left[\frac{f_1k^2}{e_1^2+f_1^2-e_1k^2} \right]+\arctan\left[
\frac{f_2k^2}{e_2^2+f_2^2-e_2k^2} \right],
\eeq
with the constraint
\beq
\eta=\tan\beta=0.0254, \qquad \beta=-\epsilon(ib). 
\eeq 
Here we took, for simplicity, the value $b=0.23135$ from our simplest fit\footnote{The experimental value for the deuteron mixing parameter $\eta=0.0254(2)$ was taken from \cite{Garcon}.} and found
\begin{equation}
E_1=e_1+if_1=-0.13304+0.004879i\quad
{\rm and} \quad E_2=e_2+if_2=-34.88+36.13i,
\end{equation}
where $E_{1,2}$ are two complex energies parametrizing
\begin{equation}
z(k)=\prod_{m=1}^2\frac{1-k^2/E_m^*}{1-k^2/E_m}.
\label{zk}
\end{equation}
This form solves the constraints (\ref{Cconstraint}) and (\ref{Ck0}) for $z(k)={\rm e}^{-2i\epsilon(k)}$.
Fig. \ref{epsilon} shows our fit for the mixing angle. We found that already 2 (complex) parameters
in (\ref{zk}) give a satisfactory fit. Using 3 parameters there is only marginal improvement.

We have 3 poles from the ${}^3S_1$ channel (\ref{Sabc}) with one of them being the bound state.
From the ${}^3D_1$ channel (\ref{B}), we get 5 poles, however only 3 of them are located
in the upper half of the complex plane. From $z(k)$ and its inverse
we get 4 poles in the upper half plane.
As a result, the Marchenko equation is transformed to a linear problem
with 9 residues which we solved algebraically.  We present our final results in Fig. \ref{multich}.  
In this and the following figures, $V_{11}$ represents the $\ell=0$, $V_{22}$ the
$\ell=2$, and $V_{12}$ the mixed component of the $2\times2$ potential matrix.

\begin{figure}
        \centering
        \begin{subfigure}[b]{0.32\textwidth}
                \centering
                \includegraphics[width=\textwidth]{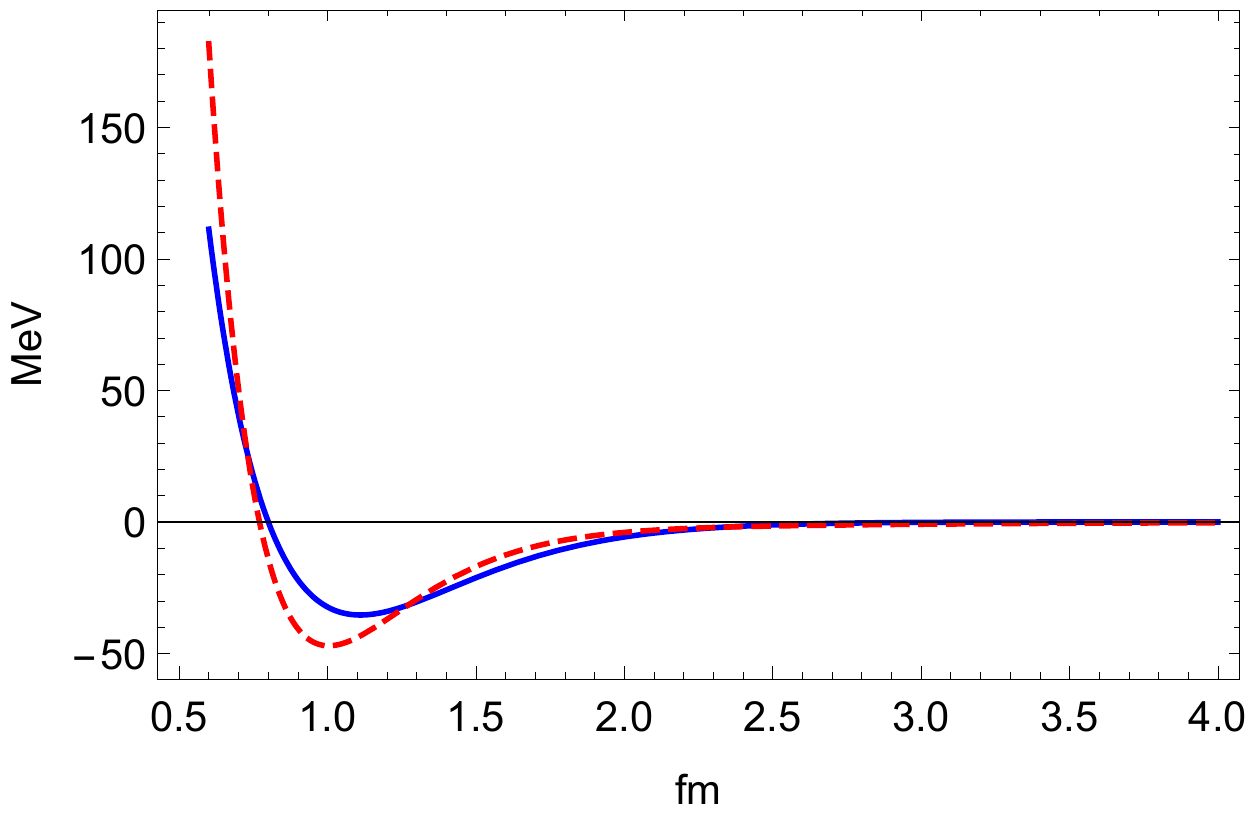}
                \caption{$V_{11}$}
                \label{V11}
        \end{subfigure}%
        ~ 
        \begin{subfigure}[b]{0.3\textwidth}
                \centering
                \includegraphics[width=\textwidth]{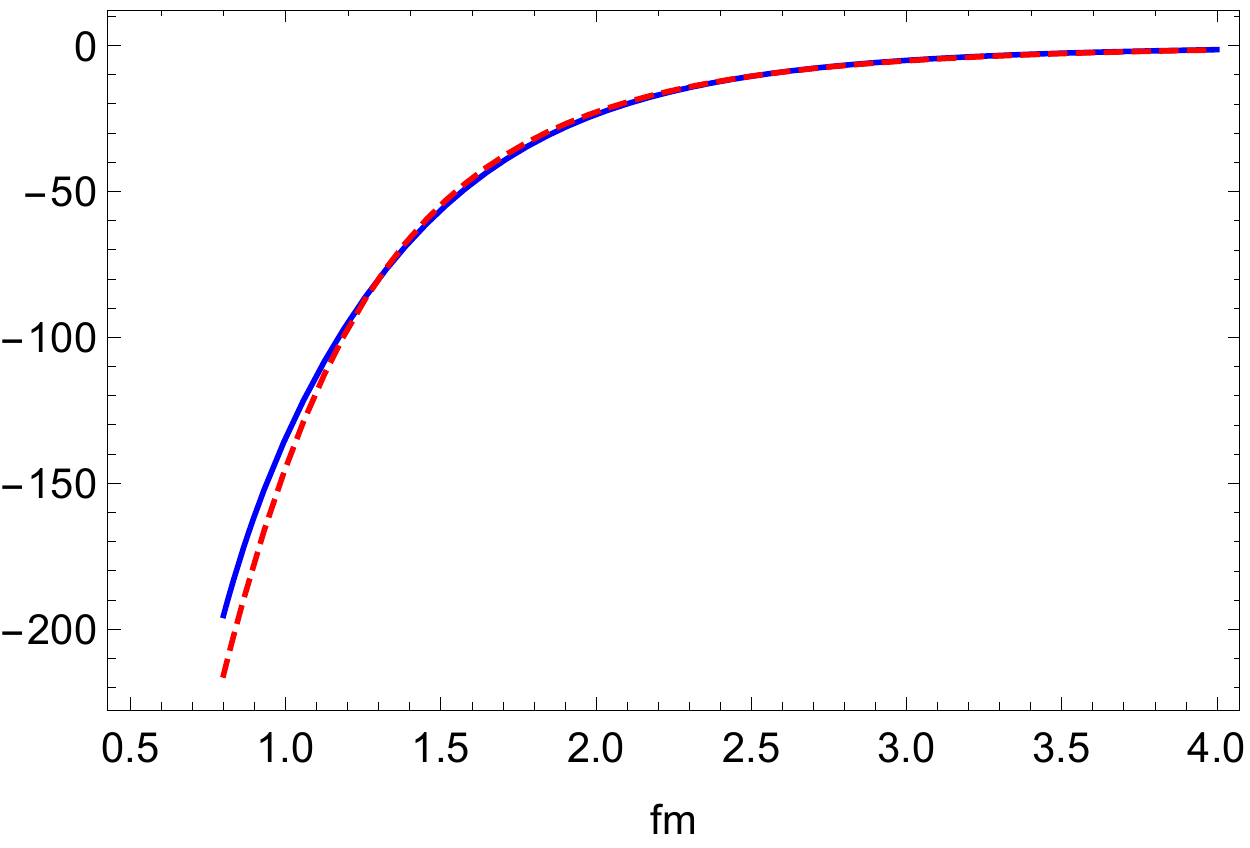}
                \caption{$V_{12}$ }
                \label{V12}
        \end{subfigure}
        ~ 
        \begin{subfigure}[b]{0.3\textwidth}
                \centering
                \includegraphics[width=\textwidth]{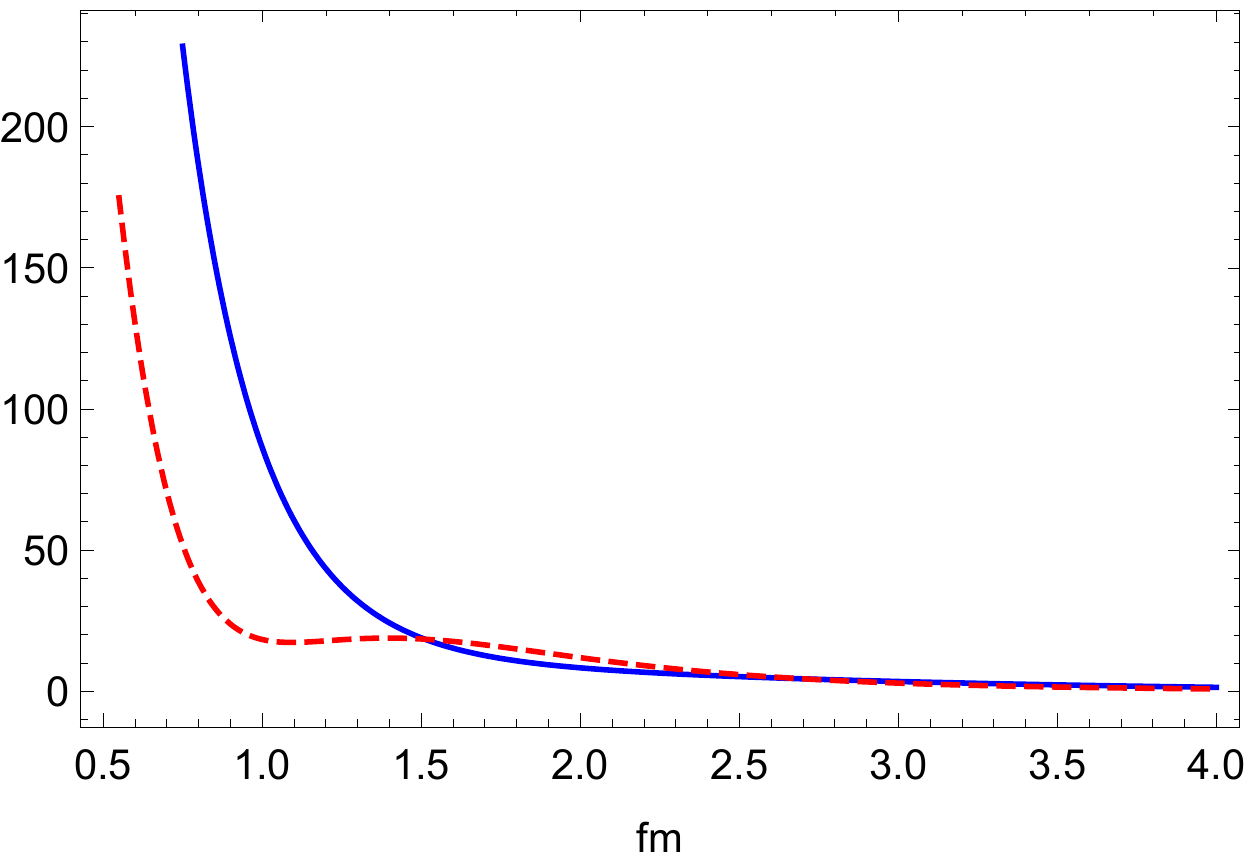}
                \caption{$V_{22}$}
                \label{V22}
        \end{subfigure}
        \caption{Matrix elements of the ${}^3S_1$-${}^3D_1$ scattering potential.
        The solid (blue) line represents our results and the dashed (red) line belongs to the  Reid93 potential.}
\label{multich}
\end{figure}

\begin{figure}
        \centering
        \begin{subfigure}[b]{0.32\textwidth}
                \centering
                \includegraphics[width=\textwidth]{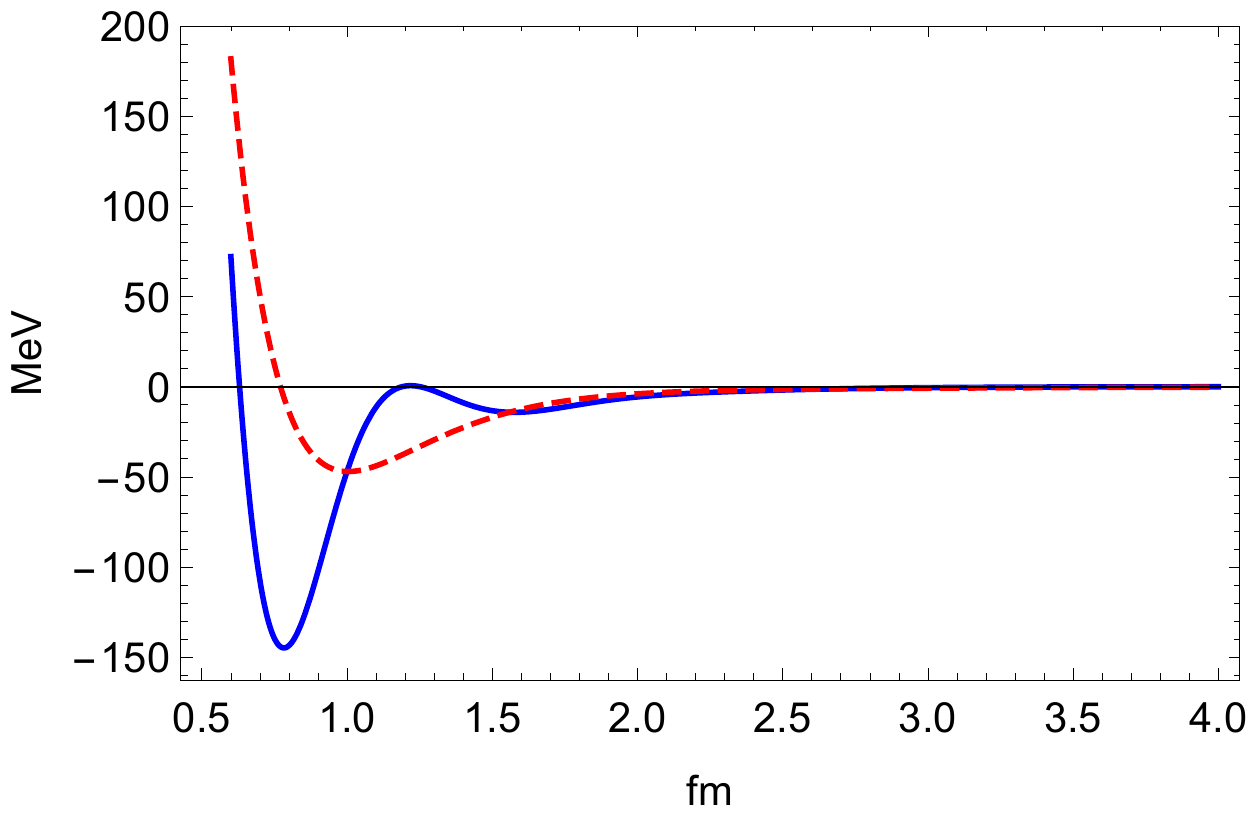}
                \caption{$V_{11}$}
                \label{V11bp}
        \end{subfigure}%
        ~ 
        \begin{subfigure}[b]{0.3\textwidth}
                \centering
                \includegraphics[width=\textwidth]{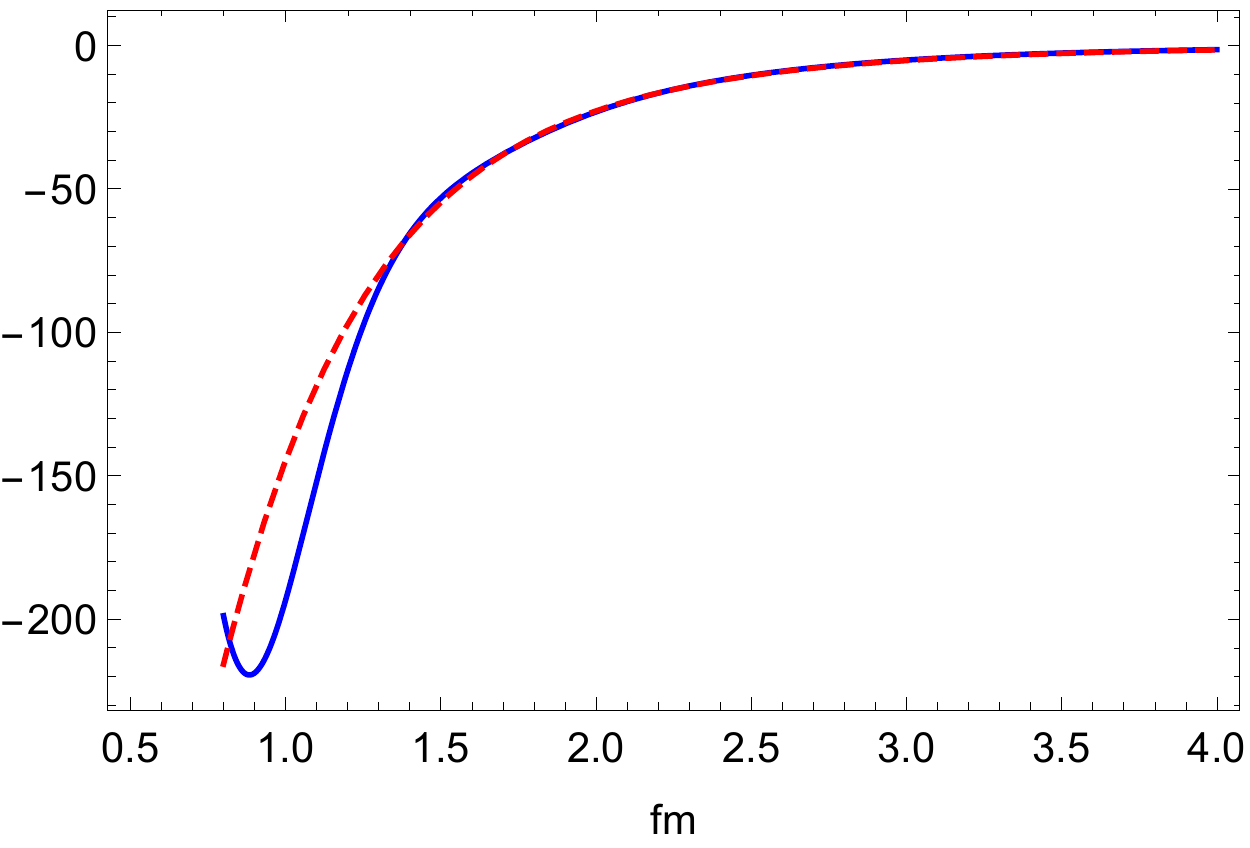}
                \caption{$V_{12}$ }
                \label{V12bp}
        \end{subfigure}
        ~ 
        \begin{subfigure}[b]{0.3\textwidth}
                \centering
                \includegraphics[width=\textwidth]{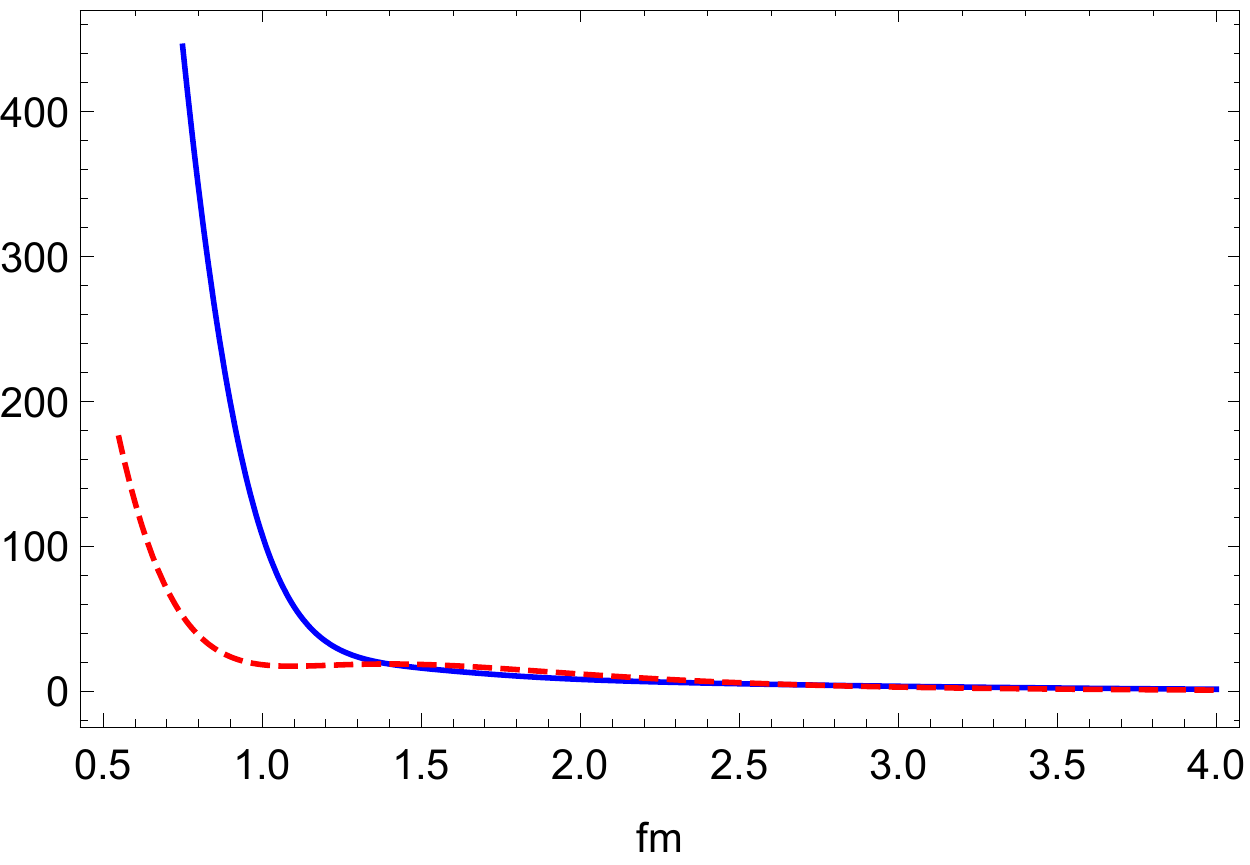}
                \caption{$V_{22}$}
                \label{V22bp}
        \end{subfigure}
        \caption{${}^3S_1$-${}^3D_1$ scattering potential obtained from the s-wave data
        in \cite{BP}. The solid (blue) line is the result of our multi-channel calculations.}        
\label{multichBP}
\end{figure}

We also studied the same problem using (\ref{Sabcd}) of Ref. \cite{BP} for the s-wave part while
keeping our d-wave parameters (\ref{B}) and the mixing angle (\ref{mixingfit}).
In this parametrization, we have 10 residues. The results are shown in Fig. \ref{multichBP}.

\begin{figure}
        \centering
        \begin{subfigure}[b]{0.32\textwidth}
                \centering
                \includegraphics[width=\textwidth]{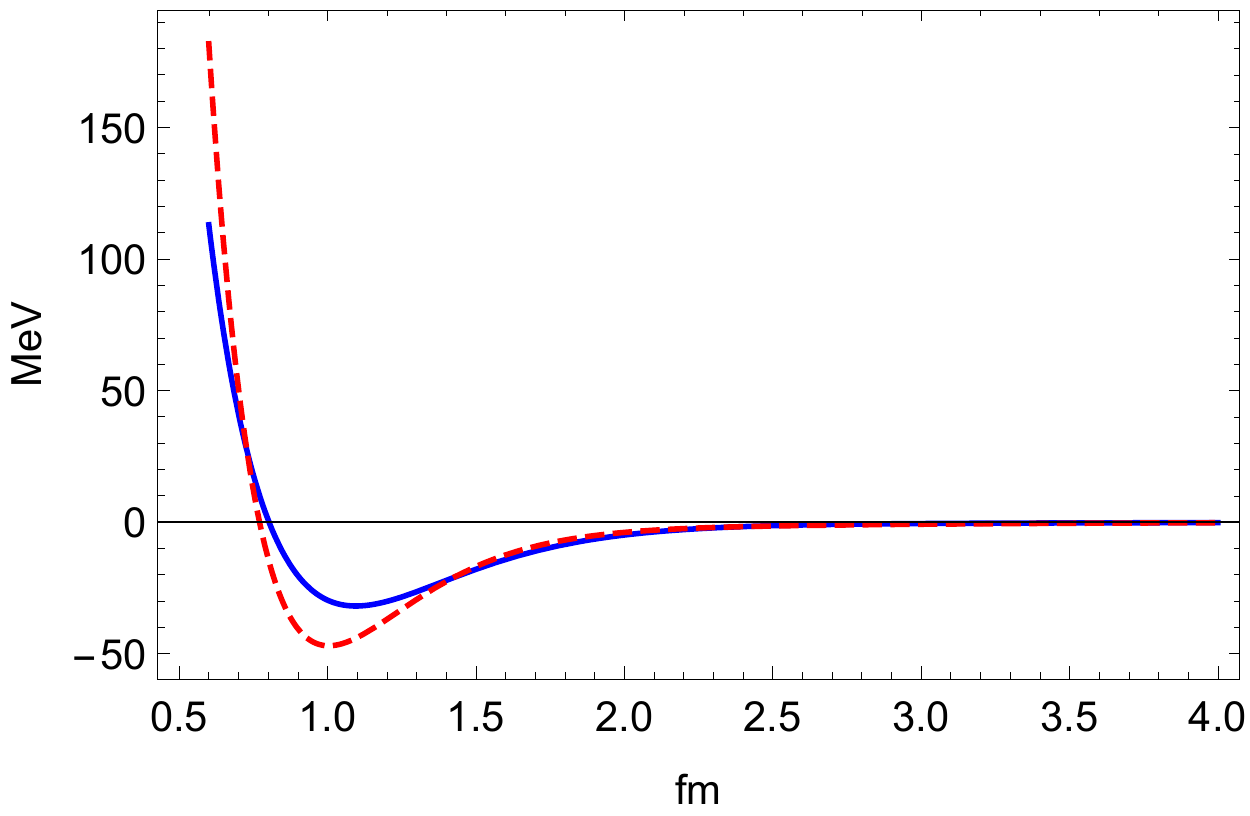}
                \caption{$V_{11}$}
                \label{V11X}
        \end{subfigure}%
        ~ 
        \begin{subfigure}[b]{0.3\textwidth}
                \centering
                \includegraphics[width=\textwidth]{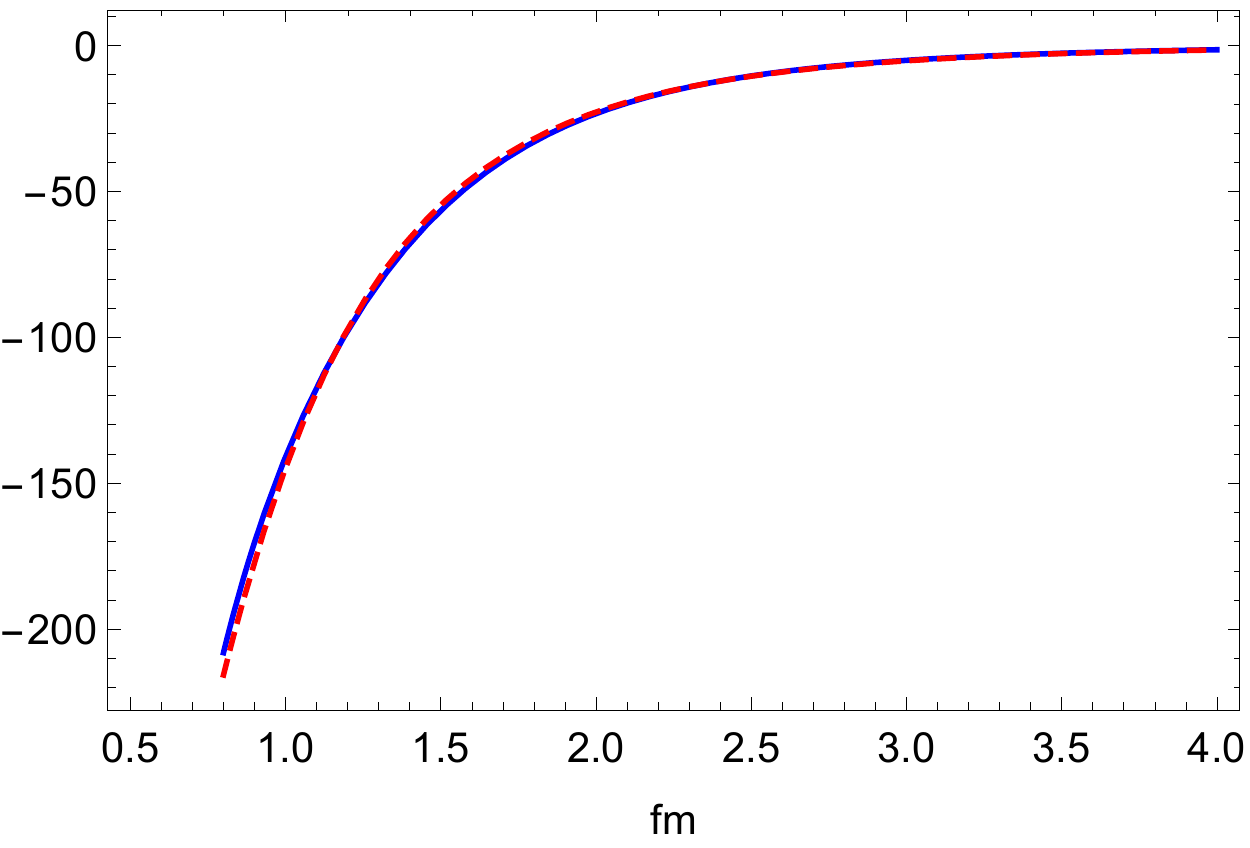}
                \caption{$V_{12}$ }
                \label{V12X}
        \end{subfigure}
        ~ 
        \begin{subfigure}[b]{0.3\textwidth}
                \centering
                \includegraphics[width=\textwidth]{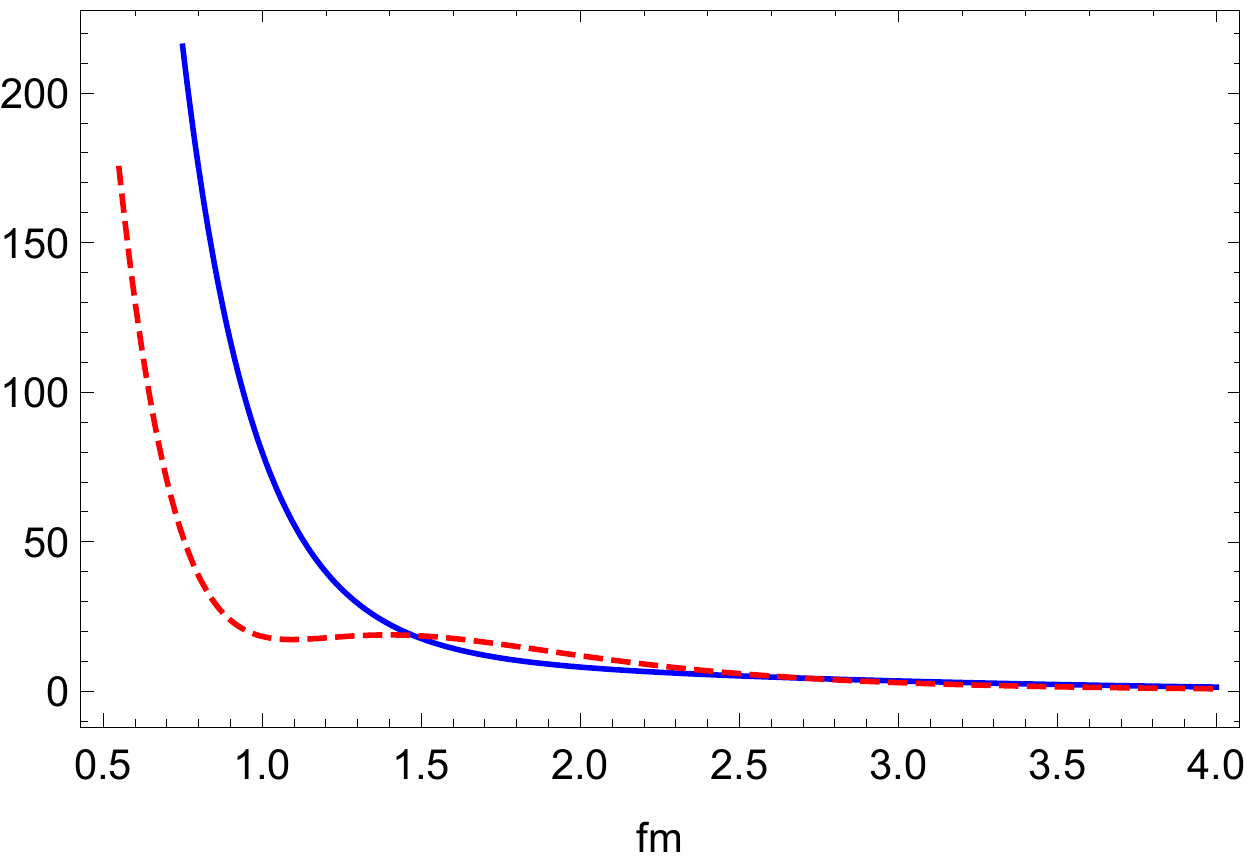}
                \caption{$V_{22}$}
                \label{V22X}
        \end{subfigure}
        \caption{${}^3S_1$-${}^3D_1$ scattering potential generated from the s-wave data in \cite{X}.}\label{multichX}
\end{figure}

Lastly, we take the s-wave roots (\ref{srootsX}) of Ref. \cite{X} and use our d-wave parameters
(\ref{B}) and the mixing angle (\ref{mixingfit}). This is, again, a
coupled problem with 10 poles. We plot these results in Fig. \ref{multichX}.

Focusing on the s-wave components of potential matrices, we see that Fig. \ref{V11} and Fig. \ref{V11X} are very similar 
and closely resemble the phenomenological potential while the solution shown in Fig. \ref{V11bp} is different, very deep and 
steeply rising for small $r$.
Our conclusion after including the mixing effects is thus unchanged: the data strongly suggest a
singular potential with strength parameter $\nu=1$ in the ${}^3S_1$ $np$ channel.

\section{${}^1S_0$ partial wave and potential}
\label{1S0}

In this section we consider the ${}^1S_0$ partial wave of the neutron-proton scattering
as a relevant physical example without multi-channel mixing. We again concentrate 
on the role of the strength of the short distance singularity $\nu$.

We begin our discussion with reviewing former studies, which were based on SUSY quantum mechanics. 

\begin{figure}
        \centering
        \begin{subfigure}[b]{0.42\textwidth}
                \centering
                \includegraphics[width=\textwidth]{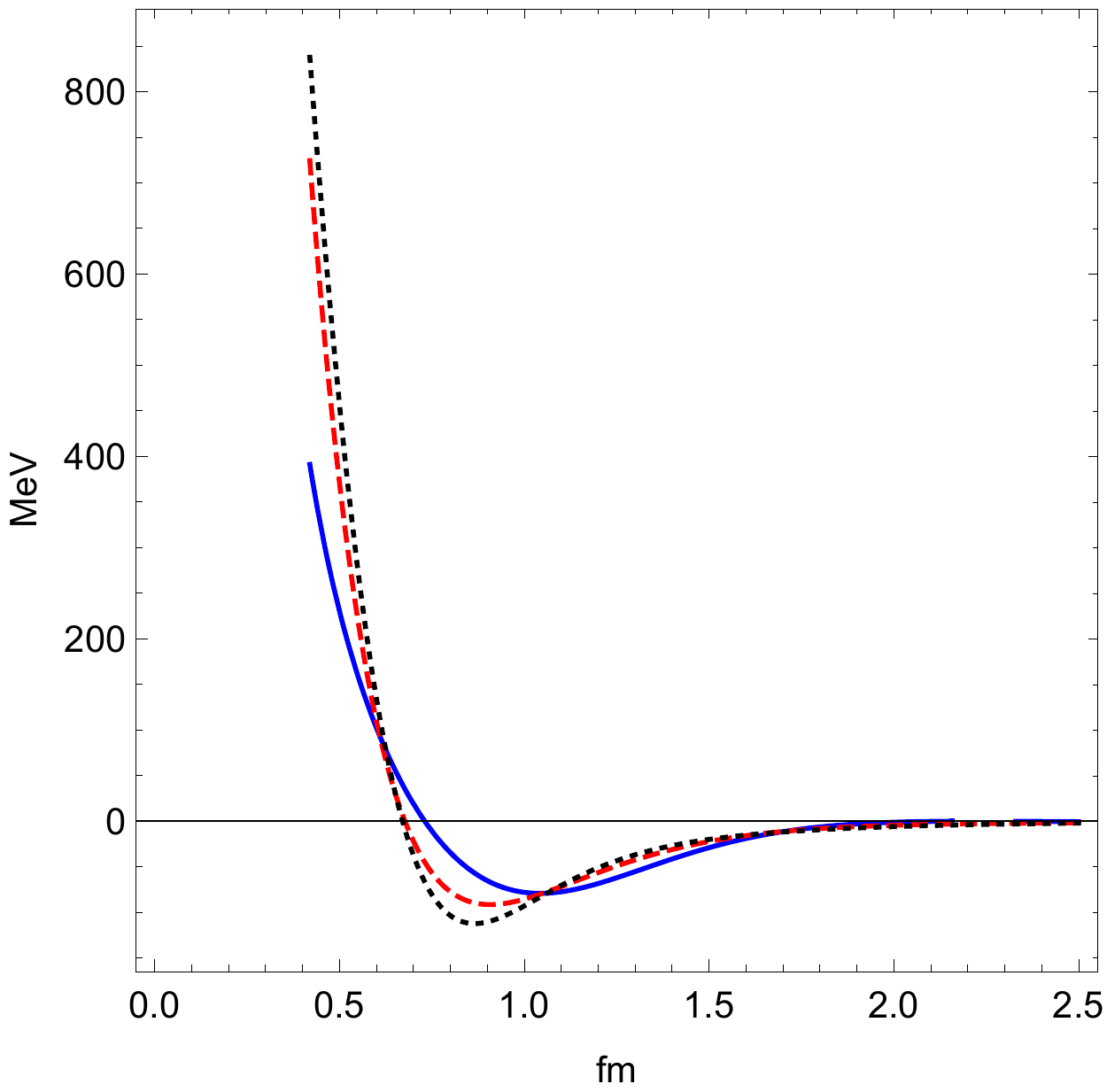}
                \caption{$\nu=1$}
                \label{susyp2}
        \end{subfigure} \quad \quad%
        ~ 
        \begin{subfigure}[b]{0.4\textwidth}
                \centering
                \includegraphics[width=\textwidth]{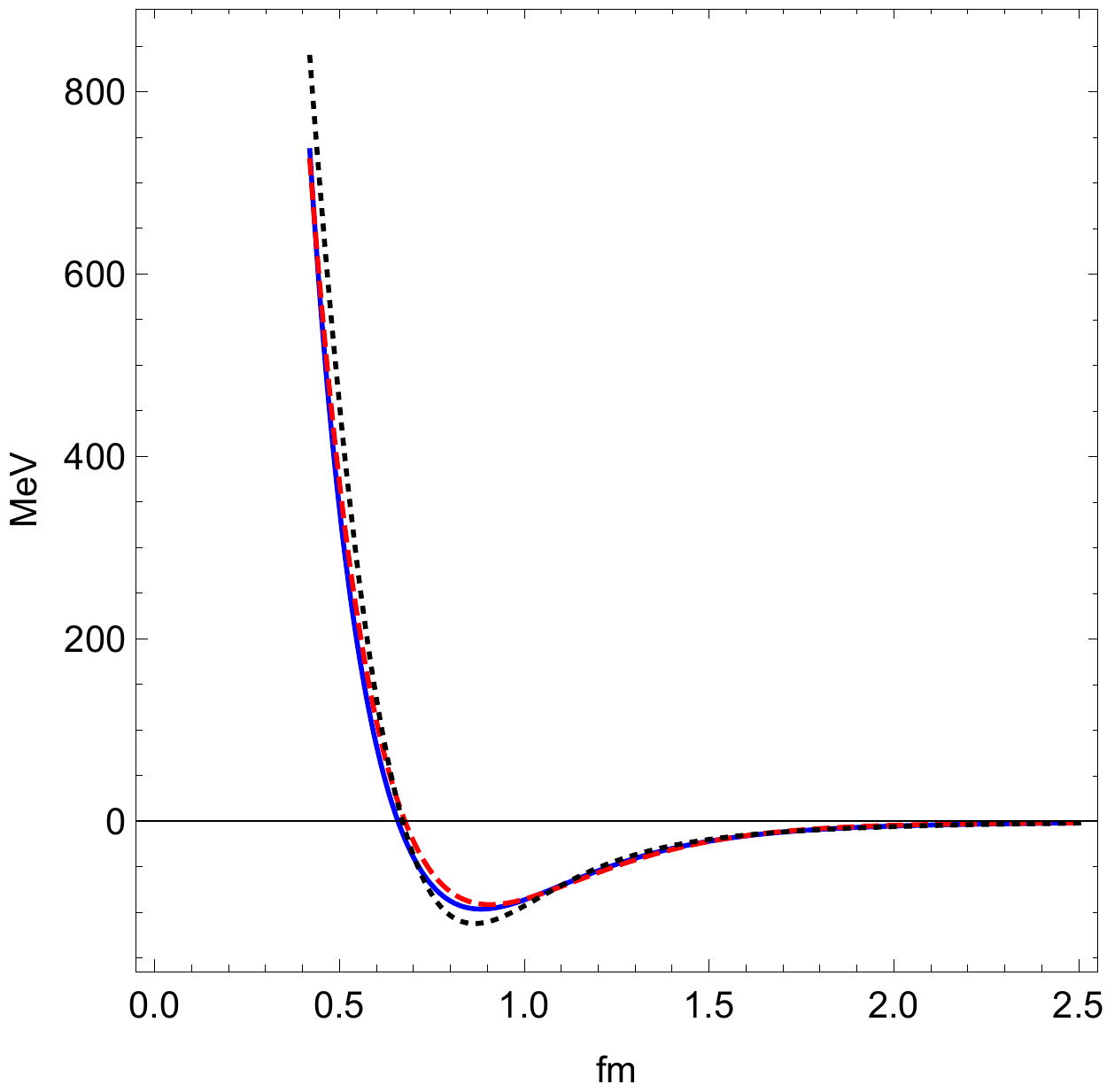}
                \caption{$\nu=2$}
                \label{susyp3}
        \end{subfigure}
        ~ 
   \label{1S0susy}
\caption{${}^1S_0$ potential. The solid (blue) lines represent the results in the cited papers,
the dashed (red) line is the Reid93 and the dotted (black) line is the AV18 reference potential.}
\end{figure}

In \cite{SparBay} the S-matrix 
\beq
S_1(k)=\sigma_a(k)\sigma_b(k)\sigma_c(k)\sigma_d(k)\sigma_e(k),
\eeq
with the roots $a=-0.040$, $b=-0.837$, $c=0.581$, $d=1.453+1.313i$, $e=d^*$
is proposed. According to Levinson's theorem (\ref{Levi}), the singularity degree is $\nu=1$.
Therefore, the resulting potential has the form $U\to\frac{2}{r^2}$ in the limit $r\to 0$.
Our calculations with the Marchenko method are presented in Fig. \ref{susyp2}. 

However, the correct short range behavior of the potential, as suggested by \cite{Stancu}, 
should be $U\to\frac{6}{r^2}$ corresponding to a singularity degree $\nu=2$. Indeed, in \cite{Stancu}
such a scattering matrix with $6$ poles is considered:
\beq
S_2(k)=\sigma_a(k)\sigma_b(k)\sigma_c(k)\sigma_d(k)\sigma_e(k)\sigma_f(k),
\eeq
$a=-0.0401$, $b=-0.7540$, $c=0.6152$, $d=2.0424$, $e=4.1650$, $f=4.6000$.
We derived the associated potential and plot it in Fig. \ref{susyp3}. As it can be seen
directly from the figure, when the singularity
degree is correct, the resulting potential has a good match with the phenomenological potential.

\begin{figure}
        \centering
        \begin{subfigure}[b]{0.42\textwidth}
                \centering
                \includegraphics[width=\textwidth]{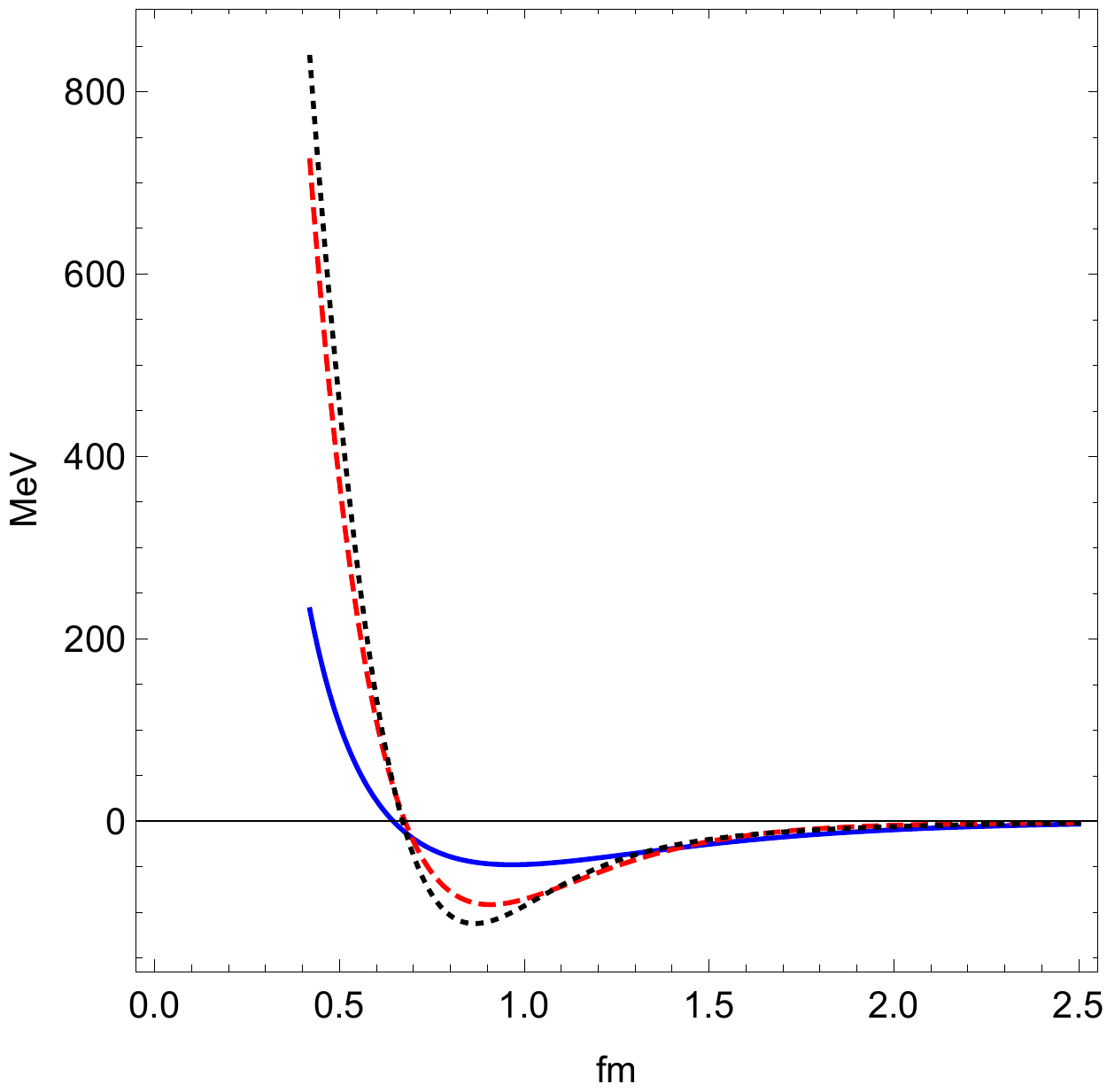}
                \caption{$\nu=1$}
                \label{marchenkop2}
        \end{subfigure} \quad \quad%
        ~ 
        \begin{subfigure}[b]{0.4\textwidth}
                \centering
                \includegraphics[width=\textwidth]{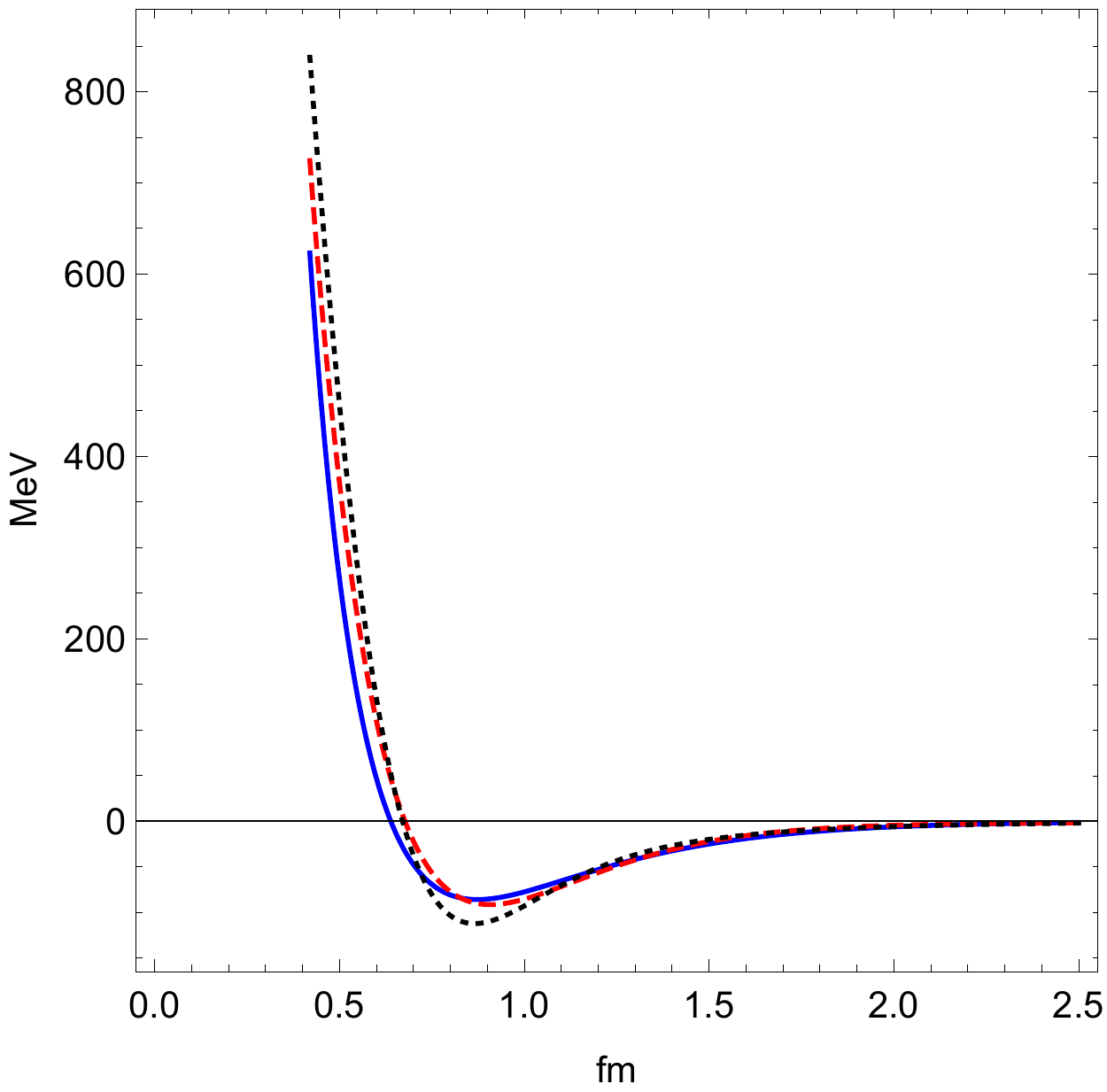}
                \caption{$\nu=2$}
                \label{marchenkop3}
        \end{subfigure}
        ~ 
   \label{1S0marchenko}
\caption{${}^1S_0$ potential. The solid (blue) lines represent our results,
the dashed (red) line is the Reid93 and the dotted (black) line is the AV18 reference potential.}
\end{figure}

We have made some new fits to the same problem. We consider first the simplest fit with
a minimal number of parameters, which we calculated analogously to the method leading to (\ref{Sabc}) and found the
following S-matrix:
\beq
S_3(k)=\sigma_a(k)\sigma_b(k)\sigma_c(k),
\eeq
with one of the poles in the lower half plane:
$a=-0.04007$, $b=1.1061$, $c=2.9254$.
According to Levinson's theorem, the related singularity degree is $\nu=1$. Here we can use simple formulas
similar to the ones in subsect. \ref{simplest1} to derive the potential, which is plotted in Fig. \ref{marchenkop2}. As clearly seen from the figure,
the potential has a wrong short range
behavior although in the long range there is a good match.  

Since the simplest 3-parameter fit is not satisfactory here, we next considered a fit with
four parameters. We found
\beq
S_4(k)=\sigma_a(k)\sigma_b(k)\sigma_c(k)\sigma_d(k),
\eeq
with $a=-0.04034$, $b=1.4561$, $c=5.2856$, $d=4.9680$.
This scattering matrix yields a potential with the correct singularity degree $\nu=2$.
We plot our result in Fig. \ref{marchenkop3}.

We see that, similarly to the ${}^3S_1$-${}^3D_1$ problem studied in sections 3 and 4, increasing the number
of fit parameters makes the shape of the potential better only if the correct singularity strength
is ensured. In the ${}^1S_0$ example discussed in this section the 3 and 5 parameter solutions both have
$\nu=1$ and disagree with the phenomenological potential for small $r$ whereas the 4 and 6 parameter
solutions (both with $\nu=2$) give nearly equally good description of the potential, including the
repulsive core.

\section{Conclusion}

In this paper we have shown that scattering data for the triplet channel
nucleon scattering strongly suggest a singular potential with inverse square
type singularity and strength $2/r^2$.
In this class of potentials already the simplest 3-parameter Bargmann-type
fit reproduces all qualitative features of the phenomenological potential.
In this case the potential and the deuteron wave function can be given in
very simple analytic form. Deuteron data are also reproduced within 1 percent
error and this representation of the phase shifts gives a maximal error of
$1.5^\circ$ in the energy range between 0 and $350\,{\rm MeV}$.

We have compared our results to the 4 and 5-parameter fits of Refs. \cite{BP}
and \cite{X} respectively. With more parameters the agreement with the phase
shifts and deuteron parameters is of course better, but while the effective
range functions ((\ref{RtBP}) and (\ref{RtX}) respectively) both look like
small refinements of (\ref{Pade11}) of the simplest fit only, there is a
drastic difference between the two cases. While the 5-parameter solution
shows no qualitative difference from our simple case, the 4-parameter solution
is very different, the potential has a very different shape, it is more
singular and steeper. The reason is that while the simplest 3-parameter
potential and the 5-parameter one both go like $2/r^2$ for short distances,
the 4-parameter one has a much stronger $6/r^2$ singularity.

Although the triplet scattering is really a coupled channel problem, we have
demonstrated that the above conclusion remains unchanged after the coupling
to the higher angular momentum d-wave and taking account of the mixing of
the two channels.

We have also studied the same questions in the singlet channel and found that
the correct singularity strength is also important there. However, the
correct singularity turns out to be $6/r^2$. This is expected since the
interaction must be more repulsive in this channel.

We have applied the Marchenko method of inverse scattering to construct the
potential and wave function from Bargmann-type rational representation of
the scattering matrix. We have shown that this method is very effective,
because the Marchenko integral equation in this case is reduced to an
algebraic problem of matrix inversion and gives compact analytical expressions
for both the potential and the wave function. Our determinant formula
(\ref{detfo}) closely resembles the analogous Crum-Krein formula used in
SUSY quantum mechanics \cite{review} but the interesting question of finding
an explicit mapping between them will be left for future studies.

\clearpage

\vspace{5ex}
\begin{center}
{\large\bf Acknowledgments}
\end{center}

We thank A. M. Shirokov for discussions.
This work was supported by 
the National Natural Science Foundation of
China (Grant No. 11575254), by the Chinese Academy of Sciences
President's International Fellowship
Initiative (Grant No. 2017PM0045 and Grant No. 2017VMA0041) and
by the Hungarian National Science Fund OTKA (under K116505). 
J.~B. would like to thank the CAS
Institute of Modern Physics, Lanzhou, where most of this work has been 
carried out, for hospitality. 

\setcounter{footnote}{0}

\par\bigskip


\appendix

\section{Scattering and inverse scattering for singular potentials}
\label{appA}

In this appendix we consider the scattering theory in the $\ell$th partial wave,
where the angular momentum quantum number $\ell$ is a nonnegative integer.
The theory of quantum inverse scattering \cite{QIS1a,QIS1b,QIS1c} in the case
of regular potentials is well established. Here we
closely follow the method of Ref. \cite{QIS2}, with modifications necessary
for potentials singular at the origin. Although in this paper all applications
are based on rational S-matrices and the corresponding singularity strength
values are integers, for completeness we discuss the general case of
arbitrary $\nu\geq0$.

We begin with the radial Schr\"{o}dinger equation
\beq
\label{radialse}
-\frac{\hbar^2}{2m}u^{\prime\prime}(r)+\frac{\hbar^2}{2m}\,
\frac{\ell(\ell+1)} {r^2}u(r)+V(r)u(r)=Eu(r),
\eeq
where $m$ is the reduced mass, $V$ is the interaction potential and
$E$ is the total energy of the particles. We introduce
\beq
q(r)=\frac{2m}{\hbar^2}V(r),\qquad k^2=\frac{2mE}{\hbar^2},
\qquad U(r)=q(r)+\frac{\ell(\ell+1)}{r^2},
\eeq
 which allows us to simplify (\ref{radialse}) as
\beq
-u^{\prime\prime}(r)+U(r)u(r)=k^2u(r).
\label{diff}
\eeq
 
We will consider potentials which are singular at the origin and vanish
exponentially at large distances,
\begin{eqnarray}
q(r)&\sim&\frac{\beta_o}{r^2}+{\rm O}(1), \qquad\quad \quad r\to0,\\
q(r)&\sim&e^{-2\kappa r},\qquad\quad r\to\infty, \quad (\kappa>0).  
\end{eqnarray}

The small $r$ singularity of the total potential term in (\ref{diff})
is determined by the parameter $\nu$ defined by
\beq
\nu(\nu+1)=\beta_o+\ell(\ell+1),
\label{beta0}
\eeq
such that
\begin{eqnarray}
U(r)&\sim&\frac{\nu(\nu+1)}{r^2}+{\rm O}(1),
\quad\qquad \qquad \  r\to0,\label{class1}\\
U(r)&\sim&\frac{\ell(\ell+1)}{r^2}+\text{exp. small},\qquad\quad r\to\infty.
\label{class2}
\end{eqnarray}


\subsection{Solutions of the direct scattering problem}

We will use two special solutions of the Schr\"{o}dinger
equation (\ref{diff}), which are specified according to their behaviour near the
origin or at infinity. 

The regular solution\footnote{Note that the generic solution is singular
at $r=0$ like $r^{-\nu}$.}
$\varphi(r,k)\equiv\varphi_k(r)$ is vanishing at $r=0$ like
\beq
\varphi_k(r)\sim D r^{\nu+1} \qquad\quad \quad r\to0,
\eeq   
where the constant $D$ is chosen for later convenience as 
\beq
D=\frac{\sqrt{\pi}}{2^{\nu+1}\Gamma(\nu+3/2)}.
\eeq
The regular solution is a real analytic and even function of $k$ for all complex $k$:
\beq
\varphi_k^*(r)=\varphi_{k^*}(r) \quad \text{and} \quad \varphi_k(r)=\varphi_{-k}(r).
\eeq 
The Jost solution $f(r,k)\equiv f_k(r)$ is determined by its asymptotic behaviour 
\beq
f_k(r)\sim e^{ikr}, \qquad\quad r\to\infty. 
\eeq
It is well-defined in the upper half plane, ${\rm{Im}}\, k\geq 0 \ (k\neq 0)$
and it is analytic for ${\rm Im}\, k> 0$. $f_k(r)$ satisfies the property
\beq
f_k^*(r)=f_{-k^*}(r).
\eeq 
The Jost function $f(k)$ is the Wronskian determinant of the regular solution
and the Jost solution
\beq
f(k)={\rm{W}}[\varphi_k, f_k]=\varphi'_k f_k-\varphi_k f'_k,
\eeq
where for any two solutions of the Schr\"odinger equation (with the same $k$)
the Wronskian is defined by 
\beq
{\rm W}[\psi_1,\psi_2]=\psi_1^\prime \psi_2-\psi_1\psi_2^\prime
\eeq
and is a constant. The Jost function is well-defined in the upper
half plane i.e. ${\rm{Im}}\,k\geq0\ (k\neq 0)$ and analytic for
${\rm{Im}}\,k>0$. It also satisfies 
\beq
f^*(k)=f(-k^*)
\eeq
and it cannot vanish for real nonzero $k$.  

For real $k$ the regular solution $\varphi_k(r)$ can be written as a linear combination
of $f_k(r)$ and $f_{-k}(r)$ as
\beq
\varphi(r,k)=\frac{i}{2k}[f(k)f_{-k}(r)-f(-k)f_k(r)], \qquad \quad (k\neq0).
\eeq
For real $k$ it is useful to introduce the phase and modulus of the
Jost function by
\beq
f(k)=\vert f(k)\vert{\rm e}^{-i\tilde \delta(k)}.
\eeq
The large $r$ asymptotics of the real solution can then be written as
\beq
\varphi_k(r)\sim \frac{\vert f(k)\vert}{k}\sin[kr+\tilde\delta(k)].
\eeq

\subsection{The $\frac{\nu(\nu+1)}{r^2}$ potential}

The solution of (\ref{diff}) with $U(r)=\frac{\nu(\nu+1)}{r^2}$ can be given
in terms of Bessel and Hankel functions. (Note that for $\nu=\ell$
this is the free case, i.e. $q(r)=0$, but for generic $\nu$ the potential
is not in our class (\ref{class1}-\ref{class2}).)
The Hankel functions are
\beq    
\qquad H_\alpha^{(1,2)}(x)=J_\alpha(x)\pm i Y_\alpha(x)
\eeq
where $J_\alpha$ and $Y_\alpha$ are the Bessel functions and we define the
rescaled Hankel functions
\beq
h_\nu(x)=\sqrt{\frac{\pi x}{2}}H_\alpha^{(1)}(x),\qquad
\tilde h_\nu(x)=\sqrt{\frac{\pi x}{2}}H_\alpha^{(2)}(x),
\eeq
where $\alpha=\nu+1/2$. Their asymptotic behaviour for large arguments is
\beq
h_\nu(x)\sim{\rm e}^{-\frac{i(\nu+1)\pi}{2}}\,{\rm e}^{ix},\qquad\quad
\tilde h_\nu(x)\sim{\rm e}^{\frac{i(\nu+1)\pi}{2}}\,{\rm e}^{-ix}
\eeq
in the upper and lower half planes, respectively.
For integer $\ell$, $h_\ell$ and $\tilde h_\ell$ are given by the finite series 
\begin{eqnarray}
h_\ell(x)&=&(-i)^{\ell+1}e^{ix}\sum^\ell_{m=0}\frac{i^m}{m!(2x)^m}
\frac{(\ell+m)!}{(\ell-m)!},\label{han}\\
\tilde{h}_\ell(x)&=&i^{\ell+1}e^{-ix}\sum^\ell_{m=0}\frac{(-i)^m}{m!(2x)^m}
\frac{(\ell+m)!}{(\ell-m)!}.
\end{eqnarray}
The regular and Jost solutions in this case are
\beq
\varphi_k(r)=k^{-\alpha}\sqrt{\frac{\pi r}{2}}J_\alpha(kr),\qquad\quad
f_k(r)={\rm e}^{\frac{i(\nu+1)\pi}{2}}h_\nu(kr)
\eeq
respectively, and so in the upper half plane (${\rm Im}\,k\geq0$)
\beq
f(k)=(-ik)^{-\nu}
\label{inf1}
\eeq
and for real $k$
\beq
\tilde\delta(k)=-\frac{\nu\pi}{2}{\rm sgn}(k).
\label{inf2}
\eeq
This is not continuous at $k=0$.

For $\nu=\ell$, $k^\ell f_k(r)$ is an entire function and 
\beq
\sigma(k)=\frac{f(-k)}{f(k)}=(-1)^\ell .
\eeq

\subsection{Analytic extension}

It can be shown \cite{QIS2} that the Jost solution also satisfies an integral
equation and with the help of this integral equation the rescaled Jost solution
\beq
{\cal{F}}(r,k)=(-ik)^\ell f(r,k)
\eeq
can be analytically extended from the upper half plane to the larger region
\break
${\rm{Im}}\,k>-\kappa$. Moreover, in a neigbourhood of the origin (small $k$)
${\cal{F}}(r,k)$ can be written as
\beq
{\cal{F}}(r,k)={\cal A}(r,k)+ik^{2\ell+1}{\cal B}(r,k),
\eeq 
where ${\cal A}(r,k)$ and ${\cal B}(r,k)$ are real analytic and even there.

${\cal F}(k)$, $A(k)$ and $B(k)$ have analogous properties, where
\beq
{\cal{F}}(k)=(-ik)^\ell f(k)=A(k)+ik^{2\ell+1}B(k).
\eeq

Finally for real $k$ we write
\beq
{\cal F}(k)=\vert{\cal F}(k)\vert{\rm e}^{-i\delta(k)}
\eeq
and (for $k>0$) we see that
\beq
\tilde\delta(k)=\delta(k)-\frac{\pi\ell}{2}
\eeq
so $\delta(k)$ measures the deviation from the free case.

$A(k)$ and $B(k)$ can be Taylor expanded for small $k$ as
\beq
A(k)=A_o+A_1k^2+\dots,\qquad\quad B(k)=B_o+B_1k^2+\dots
\eeq
It can be shown that if $A_o=0$ (${\cal F}(0)=0$) then the zero energy
regular solution is bounded. For $\ell\geq1$ it is even normalizable, so
in this case there is a genuine zero energy bound state. Since in our
applications we never ancounter zero energy bound states (nor bounded
zero energy solutions for $\ell=0$), we will assume that $A_o\not=0$
and these are absent. All our subsequent considerations are valid under
this assumption.

\subsection{Negative energy bound states}

Negative energy bound states with energy $-\kappa_j^2$ can be identified
with zeros of the Jost function on the positive imaginary axis
\beq
f(i\kappa_j)=0, \qquad \quad \kappa_j>0,\qquad j=1,\dots,J.
\eeq
In this case the regular solution $\varphi(r, i\kappa_j)\equiv \varphi_j(r)$
is real and proportional to the Jost solution $f_j(r)=f(r,i\kappa_j)$ and thus
normalizable. The normalization constant is given by
\beq
\frac{1}{s_j}=\int_0^\infty{\rm d}r\,f_j^2(r).
\eeq
The normalized wave function $\psi_j(r)$ behaves for $r\to\infty$
asymptotically as 
\beq
\psi_j(r)\sim A_j{\rm e}^{-\kappa_j r},\qquad\quad A_j=\sqrt{s_j}.
\eeq
We can classify bound states as inner and outer according to whether
$0<\kappa_j<\kappa$ or $\kappa_j\geq\kappa$.

\subsection{S-matrix}
\label{high5}

The S-matrix is defined relative to the free case:
\beq
S(k)=\frac{{\cal{F}}(-k)}{{\cal{F}}(k)}=(-1)^\ell \frac{f(-k)}{f(k)}.
\eeq
$S(k)$ is meromorphic inside the strip $-\kappa< {\rm Im}\,k<\kappa$ and
it has poles where ${\cal{F}}(k)=0$. Further it has the properties
\beq
S(-k)=\frac{1}{S(k)}, \qquad \quad  S^*(k)=S(-k^*),\qquad\quad S(0)=1.
\eeq

For $0<{\rm Im}\,k<\kappa$, poles occur only for $k=i\kappa_j$. Thus there
is a relation between inner bound states and S-matrix poles. All other
poles of the S-matrix are unrelated to the bound state structure. The
residue at an inner pole is given by the formula
\beq
S(k)\sim -i(-1)^\ell\frac{s_j}{k-i\kappa_j},\qquad\quad k\sim i\kappa_j.
\eeq

For real $k$
\beq
S(k)=e^{2i\delta(k)},
\eeq
where $\delta(k)$ is the phase shift.
For small $k$
\beq
S(k)=\frac{A(k)-ik^{2\ell+1}B(k)}{A(k)+ik^{2\ell+1}B(k)}\sim 1-2ik^{2\ell+1}\frac{B_o}{A_o}+\dots
\eeq
Thus, for small $k$, both $S(k)-1$ and $\delta(k)$ are O$(k^{2\ell+1})$.

\subsection{Levinson's theorem}

One of the crutial modifications of the theory if the potential
is singular occurs in Levinson's theorem \cite{Swan}:
\beq
\label{Levi}
\delta(\infty)-\delta(0)=\int_0^\infty{\rm d}k\,\delta^\prime(k)=-\frac{\pi}{2}(2J+\nu-\ell).
\eeq 
This is consistent with (\ref{inf1}-\ref{inf2}) since in the $k\to\infty$ limit
the solutions approach that of the $\nu(\nu+1)/r^2$ potential.

\subsection{Inverse scattering}

Starting from the completeness relation
\beq
\frac{2}{\pi}\int_0^\infty\frac{k^2{\rm d}k}{\vert f(k)\vert^2}
\varphi(r,k)\varphi(s,k) + \sum_{j=1}^J s_j f_j(r) f_j(s)=\delta(r-s)
\eeq
and following the derivation in \cite{QIS2} we obtain, for the
unknown $A(r,s)$, the following Marchenko equation
\beq
F(r,s)+A(r,s)+\int_r^\infty{\rm d}u\,A(r,u)F(u,s)=0, \qquad s\geq r.
\label{Marchenko}
\eeq
Here
\beq
F(r,s)=F_{\rm bound}(r,s)+F_{\rm scatt}(r,s)
\label{Fbs}
\eeq 
and
\beq
F_{\rm bound}(r,s)=(-1)^{\ell+1}\sum_{j=1}^J s_j\
h_\ell(i\kappa_jr)\ h_\ell (i\kappa_js),
\label{Fb}
\eeq
\beq
\begin{split}
&F_{\rm scatt}(r,s)=-\frac{\sin\nu\pi}{\pi(r+s)}+\frac{(-1)^{\ell+1}}{2\pi}
\int_0^\infty{\rm d}k\,\Big\{[S(k)-1]y(r,k)y(s,k)\\
&+[S(-k)-1]y(r,-k)y(s,-k)+[1-S(\infty)]{\rm e}^{ik(r+s)}
+[1-S(-\infty)]{\rm e}^{-ik(r+s)}\Big\}
\end{split}
\label{Fs}
\eeq
with
\beq
y(r,k)=i^{\ell+1}\,h_\ell(kr).
\eeq
Note that for the simplest case of $\ell=0$ and non-singular potential
($\nu=0$) the usual formulas
\beq
F_{\rm bound}(r,s)=\sum_{j=1}^Js_j{\rm e}^{-\kappa_j(r+s)}
\eeq
and
\beq
F_{\rm scatt}(r,s)=\frac{1}{2\pi}\int_{-\infty}^\infty{\rm d}k\,
[1-S(k)]{\rm e}^{ik(r+s)}
\label{Fs2}
\eeq
are reproduced. For integer $\nu$ (\ref{Fs}) can be rewritten as
\beq
F_{\rm scatt}(r,s)=\frac{1}{2\pi}\int_{-\infty}^\infty{\rm d}k\,\big\{[S(k)-1]
h_\ell(rk)h_\ell(sk)+[(-1)^\nu-(-1)^\ell]{\rm e}^{ik(r+s)}\big\}.
\label{intnu}
\eeq

If we can calculate the $F(r,s)$ function from scattering data and
bound state information and can solve the Marchenko equation
(\ref{Marchenko}) for $A(r,s)$ then not only the potential can be
expressed as
\beq
q(r)=-2\frac{{\rm d}}{{\rm d}r}\,A(r,r),\qquad\quad U(r)=q(r)
+\frac{\ell(\ell+1)}{r^2}
\eeq
but also the wave function is calculable \cite{QIS2}. It is given
by the formula
\beq
f(r,k)=y(r,k)+\int_r^\infty{\rm d}u\,A(r,u)y(u,k).
\eeq.


\section{Marchenko method for Bargmann-type S-matrices}
\label{appB}

In this appendix we apply the Marchenko method of inverse scattering to
Bargmann-type (rational) S-matrices.

\subsection{Rational S-matrices}

A general rational solution of the $S(0)=1$, $S(k)S(-k)=1$ and $S^*(k)=S(-k^*)$
requirements is of the form
\beq
S(k)=\prod_{\alpha=1}^{{\cal N}_+}\frac{\lambda_\alpha-ik}{\lambda_\alpha+ik}
\prod_{\beta=1}^{{\cal N}_-}\frac{\mu_\beta-ik}{\mu_\beta+ik},
\label{BargS}
\eeq
where the parameters $\lambda_\alpha$ have positive real parts and they
are either real or come in complex conjugate pairs. Similarly the real
parts of $\mu_\beta$ are negative and they are also either real or form
complex conjugate pairs. We assume, for simplicity, that they are all
distinct so that the S-matrix (\ref{BargS}) has simple poles only.

The residues at the poles in the upper half plane are given by
\beq
S(k)\sim \frac{-iR_\alpha}{k-i\lambda_\alpha}\qquad k\sim i\lambda_\alpha,
\eeq
where
\beq
R_\alpha=2\lambda_\alpha
\prod_{\alpha^\prime \not=\alpha} \frac{\lambda_{\alpha^\prime}+\lambda_\alpha}
{\lambda_{\alpha^\prime}-\lambda_\alpha}
\prod_\beta \frac{\mu_\beta+\lambda_\alpha}
{\mu_\beta-\lambda_\alpha}.
\eeq
$R_\alpha$ is real for $\lambda_\alpha$ real and $R_\alpha^*=R_{\bar\alpha}$
for $\lambda_\alpha=\lambda^*_{\bar\alpha}$ complex conjugate pairs.

Given the S-matrix (\ref{BargS}) we still have to specify
$\{s_j,\kappa_j\}_{j=1}^J,\quad s_j>0,\kappa_j>0$, i.e. the number of bound
states and the corresponding binding energies and asymptotic decay constants.
Moreover, these data have to satisfy the following constraints:
\begin{itemize}

\item
For higher partial waves $\ell\geq1$ the S-matrix parameters have to be chosen
such that (\ref{BargS}) satisfies the 
\beq
S(k)=1+{\rm O}(k^{2\ell+1})
\eeq
constraint for small $k$.

\item
The singularity strength, obtained from Levinson's theorem (\ref{Levi}), is
given by
\beq
\nu={\cal N}_+-{\cal N}_--2J+\ell
\eeq
in this case and has to be a non-negative integer.

\item
Because of the relation between S-matrix poles and inner bound states
the sets
\beq
\{\lambda_\alpha\,\vert\, {\rm Re}(\lambda_\alpha)<\kappa\}\qquad
{\rm and}\qquad\{\kappa_j<\kappa\} 
\eeq
must coincide, moreover, the corresponding residues (for
$\lambda_\alpha=\kappa_j$) are restricted by $R_\alpha=(-1)^\ell s_j$, which
is possible only if
\beq
(-1)^\ell R_\alpha>0.
\eeq

\end{itemize}

The advantage of using rational S-matrices is that the singularity
strength $\nu$ is always an integer and when calculating the scattering
contribution to Marchenko's $F$ function we can use (\ref{intnu}), which is
now easily evaluated using Cauchy's theorem by closing the contour in the
upper half plane:
\beq
F_{\rm scatt}(r,s)=
\sum_\alpha R_\alpha h_\ell(i\lambda_\alpha r)
h_\ell(i\lambda_\alpha s).
\eeq
The contributions coming from the terms with ${\rm Re}(\lambda_\alpha)<\kappa$
are canceled by the same, but opposite sign contributions of the inner bound
states to $F_{\rm bound}$, so the final result for the full $F$ function is
given by
\beq
F(r,s)=
\sum_{{\rm Re}(\lambda_\alpha)\geq\kappa} R_\alpha h_\ell(i\lambda_\alpha r)
h_\ell(i\lambda_\alpha s)
+(-1)^{\ell+1}\sum_{\kappa_j\geq\kappa} s_j h_\ell(i\kappa_j r)
h_\ell(i\kappa_j s).
\eeq
Since the form of these remaining terms are the same, we can combine them
and write
\beq
F(r,s)=
\sum_{n=1}^N R_n h_\ell(ig_n r)h_\ell(ig_n s),\qquad {\rm Re}(g_n)>0,
\label{BargF}
\eeq
where the parameters $g_n$ are either real or come in complex conjugate
pairs and the coefficient $R_n$ is real for real $g_n$ and $R^*_n=R_{\bar n}$
for complex conjugate pairs.

\subsection{Algebraic equation}

For rational S-matrices the solution of the Marchenko integral equation
is completely algebraic \cite{vonGeramb1,vonGeramb2}. We have simplified
the derivation and expressed the final result in a compact form.

Let us introduce the shorthand notation $\omega_n(r)=h_\ell(ig_n r)$. Then the
Marchenko integral equation with $F(r,s)$ given by (\ref{BargF}) becomes
\beq
\sum_{n=1}^N R_n\omega_n(r)\omega_n(s)+A(r,s)+\sum_{n=1}^N R_n\omega_n(s)
\int_r^\infty{\rm d}u\,A(r,u)\omega_n(u)=0.
\eeq
Thus the solution for $A(r,s)$ has to be of the form
\beq
A(r,s)=-\sum_{n=1}^N b_n(r)\omega_n(s),
\eeq
where
\beq
\begin{split}
b_n(r)&=R_n\omega_n(r)+R_n\int_r^\infty{\rm d}u\,A(r,u)\omega_n(u)\\
&=R_n\omega_n(r)-R_n\sum_{m=1}^N b_m(r)\int_r^\infty\omega_n(u)
\omega_m(u){\rm d}u.
\end{split}
\eeq
The second equality is a linear algebraic set of equations for the unknown
coefficients $b_n(r)$:
\beq
\frac{1}{R_n}b_n(r)+\sum_{m=1}^N I_{nm}(r)b_m(r)=\omega_n(r),
\eeq
where
\beq
I_{nm}(r)
=\int_r^\infty\omega_n(u)\omega_m(u){\rm d}u
=\int_r^\infty h_\ell(ig_n u)h_\ell(ig_mu){\rm d}u.
\eeq
Using the properties of the Hankel functions, the integral of the
product of two Hankel functions can be done in terms of other Hankel
funcions\footnote{We define $h_{-1}(x)={\rm e}^{ix}$.}:
\beq
I_{nm}(r)=\frac{i}{g_n^2-g_m^2}\left\{g_mh_\ell(ig_nr)h_{\ell-1}(ig_mr)
-g_nh_\ell(ig_mr)h_{\ell-1}(ig_nr)\right\}
\label{Hint}
\eeq
for $n\not=m$ and
\beq
I_{nn}(r)=\frac{r}{2}\left\{h_{\ell-1}(ig_nr)h_{\ell+1}(ig_nr)
-h_\ell^2(ig_nr)\right\}.
\label{Hint2}
\eeq
Introducing the matrix of the linear problem,
\beq
{\cal M}_{nm}(r)=\frac{1}{R_n}\delta_{nm}+I_{nm}(r),
\eeq
we have to solve
\beq
\sum_{m=1}^N{\cal M}_{nm}(r)b_m(r)=\omega_n(r).
\eeq
The solution is
\beq
b_n(r)=\sum_{m=1}^N{\cal N}_{nm}(r)\omega_m(r),
\eeq
where ${\cal N}$ is the matrix inverse of ${\cal M}$.
To calculate the potential we only need
\beq
A(r,r)=-\sum_{n=1}^Nb_n(r)\omega_n(r)=-\sum_{n,m=1}^N\omega_n(r){\cal N}_{nm}(r)
\omega_m(r),
\eeq
which can also be written as
\beq
A(r,r)=\sum_{n,m=1}^N{\cal N}_{nm}(r)\frac{\rm d}{{\rm d}r}{\cal M}_{mn}(r)=
{\rm Tr}[{\cal M}^{-1}\frac{\rm d}{{\rm d}r}{\cal M}(r)]=
\frac{\rm d}{{\rm d}r}{\rm ln\,det}({\cal M}).
\eeq
The final result is
\beq
q(r)=-2\frac{{\rm d}^2}{{\rm d}r^2}{\rm ln\,det}({\cal M}).
\label{detfo}
\eeq

\subsection{s-wave scattering}
\label{swave}

The formulas simplify in the case of s-wave scattering ($\ell=0$). In this
case the Hankel functions become simple exponentials and Marchenko's $F$
takes the form
\beq
F(r,s)=F(x)=-\sum_{m=1}^N\,R_m\,{\rm e}^{-\omega_m x},\qquad x=r+s.
\label{expF}
\eeq
To express the solution in simple terms it is useful to define
\beq
{\cal K}_{mn}=\frac{1}{z_m}\delta_{mn}-\frac{1}{\omega_m+\omega_n},\qquad
\qquad z_m(r)=R_m\,{\rm e}^{-2\omega_mr}.
\eeq
Then
\beq
q(r)=-2\frac{{\rm d}}{{\rm d}r}A(r,r)=
-2\frac{{\rm d}^2}{{\rm d}r^2}\ln\vert{\rm det}({\cal K})\vert
\label{Axx}
\eeq
and the bound state wave functions can also be calculated algebraically:
\beq
f_j(r)={\rm e}^{-\kappa_j r}\left\{1+\sum_{m=1}^N\frac{\gamma_m(r)}{\kappa_j+\omega_m}\right\},
\label{wf}
\eeq
where
\beq
\gamma_m=\sum_{k=1}^N\left({\cal K}^{-1}\right)_{mk}.
\eeq

\subsection{Examples}
\label{B2}

In this subsection and the next we will discuss examples most of which we
used in the main text. All our examples correspond to s-wave scattering
($\ell=0$). We will use the $({\cal N}_+,{\cal N}_-)[J]$ notation and first
consider the $(3,0)[1]$ case. In this case
\beq
S(k)=\sigma_a(k)\sigma_b(k)\sigma_c(k),\qquad a,b,c >0.
\eeq
We define the constants
\beq
\tilde R_a=2a\frac{(b+a)(c+a)}{(b-a)(c-a)},\qquad
\tilde R_b=2b\frac{(a+b)(c+b)}{(a-b)(c-b)},\qquad
\tilde R_c=2c\frac{(a+c)(b+c)}{(a-c)(b-c)}.
\eeq
We will assume that the single bound state coincides with one of the parameters,
$\kappa_o=b$. The corresponding norming constant can be parametrized as
$s_o=y\tilde R_b$ and we write
\beq
F(r)=(y-1)\tilde R_b{\rm e}^{-br}-\tilde R_a{\rm e}^{-ar}-\tilde R_c{\rm e}^{-cr}.
\eeq
We will study two cases: $y=1$ and $y=2$. These are only possible if $b$ is
the smallest (or largest) parameter of the S-matrix since the norming
constant must be positive. The first choice corresponds to treating $k=ib$
as inner pole, in which case this term is eliminated from $F(r)$ completely.
The second one can be called the sign-flip choice.

In the first case where $y=1$ and $N=2$  we have
\beq
\omega_1=a, \quad \omega_2=c, \qquad\qquad 
R_1=\tilde R_a,\quad R_2=\tilde R_c.
\eeq 
 Therefore, this is a  $2\times2$ matrix problem with
\beq
{\cal K}_1=
\begin{pmatrix}
 \frac{{\rm e}^{2ar}}{\tilde R_a}-\frac{1}{2a} &-\frac{1}{a+c}  \\
-\frac{1}{a+c} & \frac{{\rm e}^{2cr}}{\tilde R_c}-\frac{1}{2c}
\end{pmatrix}
\eeq
and can easily be solved
explicitly for general $a$, $b$, $c$. The solution is given in
sect.~\ref{simplest1}.

In the sign-flip case $N=3$ and
\beq
\omega_1=b,\quad \omega_2=a,\quad \omega_3=c,\qquad\qquad R_1=-\tilde R_b,\quad
R_2=\tilde R_a,\quad R_3=\tilde R_c.
\eeq
Furthermore we take $b=1$, $a=2$ and $c=4$. The corresponding $3\times 3$
matrix is
\beq
{\cal K}_2=
\begin{pmatrix}
-\frac{1}{10}{\rm e}^{2r}-\frac{1}{2}&-\frac{1}{3}&-\frac{1}{5}\\
-\frac{1}{3}&-\frac{1}{36}{\rm e}^{4r}-\frac{1}{4}&-\frac{1}{6}\\
-\frac{1}{5}&-\frac{1}{6}&\frac{1}{40}{\rm e}^{8r}-\frac{1}{8}
\end{pmatrix}.
\eeq
From (\ref{Axx}) and (\ref{wf}) we get
\beq
q(r)=\frac{2}{\sinh^2 r}-\frac{12}{\cosh^2 r}
\eeq
and
\beq
f_o(r)=\frac{1}{2}\frac{\sinh^2 r}{\cosh^3 r}=\frac{1}{2}\varphi_o(r).
\eeq
These are the generalized P\"oschl-Teller (PT) potential and its ground state
wave function \cite{PT}.

\subsection{More examples}
\label{B3}

Let us consider more examples. Firstly, we investigate the $(4,0)[1]$ case where the scattering matrix is represented as
\beq
S(k)=\sigma_a(k)\sigma_b(k)\sigma_c(k)\sigma_d(k).
\eeq
Without going any further, we observe that $S(k)$ falls into the class of PT type potentials with the degree of singularity $\nu=2$ according to Levinson's theorem.

As before, we identify the bound state momentum $\kappa_o$ with $b$. We find 
\beq
F(r)=(y-1)\tilde R_b{\rm e}^{-br}-\tilde R_a{\rm e}^{-ar}-\tilde R_c{\rm e}^{-cr}-\tilde R_d{\rm e}^{-dr},
\eeq  
where the $\tilde R_m$-s read
\beq
\begin{split}
  \tilde R_a&=-2a\frac{(a+b)(a+c)(a+d)}{(a-b)(a-c)(a-d)},\\
  \tilde R_c&=-2c\frac{(c+a)(c+b)(c+d)}{(c-a)(c-b)(c-d)},
\end{split}\qquad
\begin{split}
  \tilde R_b&=-2b\frac{(b+a)(b+c)(b+d)}{(b-a)(b-c)(b-d)},\\ 
  \tilde R_d&=-2d\frac{(d+a)(d+b)(d+c)}{(d-a)(d-b)(d-c)}.
\end{split}
\eeq

In subsection \ref{simplest2} we used the complex parameters of Ref. \cite{BP}.
We note that this choice of the parameters yields, once more, a real and positive $\tilde R_b$.
Therefore, we are able to cancel it by choosing $y=1$. We are left with
\beq
\omega_1=a, \quad \omega_2=c, \quad \omega_3=d, \qquad\qquad 
R_1=\tilde R_a,\quad R_2=\tilde R_c, \quad R_3=\tilde R_d.\quad    
\eeq  
The corresponding potential is shown in Fig. \ref{3S1BP} (section \ref{simplest}).

Finally, we consider a $(4,1)[1]$ type S-matrix composed of 5 factors
\beq
S(k)=\sigma_a(k)\sigma_b(k)\sigma_c(k)\sigma_d(k)\sigma_e(k),
\eeq
with one of its poles ($a$) being negative \cite{X}.
Once again, the bound state momentum is $\kappa_o=b$. Since the contour is closed in upper half plane,
the negative pole does not contribute to the integral (\ref{Fs2}). Therefore we obtain
\beq
F(r)=(y-1)\tilde R_b{\rm e}^{-br}-\tilde R_c{\rm e}^{-cr}-\tilde R_d{\rm e}^{-dr}-\tilde R_e{\rm e}^{-er},
\eeq  
where 
\beq
\begin{split}
\tilde R_b&=2b\frac{(b+a)(b+c)(b+d)(b+e)}{(b-a)(b-c)(b-d)(b-e)},\\
\tilde R_d&=2d\frac{(d+a)(d+b)(d+c)(d+e)}{(d-a)(d-b)(d-c)(d-e)},
\end{split}\quad
\begin{split}
\tilde R_c&=2c\frac{(c+a)(c+b)(c+d)(c+e)}{(c-a)(c-b)(c-d)(c-e)},\\
\tilde R_e&=2e\frac{(e+a)(e+b)(e+c)(e+d)}{(e-a)(e-b)(e-c)(e-d)}.
\end{split}
\eeq
We again focus on the $y=1$ case and write 
\begin{equation}
\omega_1=c, \quad \omega_2=d, \quad \omega_3=e,\qquad\qquad
R_1=\tilde R_c, \quad R_2=\tilde R_d, \quad R_3=\tilde R_e.
\end{equation}
The corresponding potential is shown in Fig. \ref{3S1X} (section \ref{simplest}).

\section{Coupled channel problems}
\label{appC}

For coupled channel problems different partial waves $\psi_a(r)$ $a=1,...,N$
(with orbital angular momentum $\ell_a$) are mixed by the potential term
in the radial Schr\"odinger equation:
\beq
-\frac{\hbar^2}{2m}\psi_a^{\prime\prime}(r)+\sum_{b=1}^N U_{ab}(r)\,\psi_b(r)+
\frac{\hbar^2}{2m}\frac{\ell_a(\ell_a+1)}{r^2}\psi_a(r)=E\psi_a(r),
\eeq
where $m$ is the reduced mass of the system. We will assume that the
potential matrix $U_{ab}$ is real and symmetric. In the case of nucleon
scattering it is usually asumed that in addition to the total angular
momentum $j$ the total spin $S$ is also conserved, so in this application
spin triplet ($S=1$) scattering is a coupled channel problem with $N=2$ and
$\ell_{1,2}=j\mp1$.

After rescaling
\beq
U_{ab}(r)=\frac{\hbar^2}{2m} q_{ab}(r),\qquad\qquad
E=\frac{\hbar^2k^2}{2m}
\eeq
the Schr\"odinger equation becomes
\beq
-\psi_a^{\prime\prime}(r)+V_{ab}(r)\psi_b(r)=k^2\psi_a(r)
\label{coup}
\eeq
with
\beq
V_{ab}(r)=q_{ab}(r)+\frac{\ell_a(\ell_a+1)}{r^2}\delta_{ab}.
\eeq

The details of scattering and inverse scattering theory depend on the
class of potentials. Here we will consider the case of real symmetric
potentials with asymptotic behaviour
\begin{eqnarray}
q_{ab}(r)&\sim&\frac{\beta_{ab}}{r^2}+{\rm O}(1), \qquad r\to0,
\qquad\quad \beta_{ab}=\beta_{ba}\quad{\rm real},\\
q_{ab}(r)&\sim&e^{-2\kappa r},\qquad r\to\infty, \quad (\kappa>0).  
\end{eqnarray}
For the total potential term in the radial Schr\"odinger equation this implies
\begin{eqnarray}
V_{ab}(r)&\sim&\frac{\nu_{ab}}{r^2}+{\rm O}(1),
\qquad   r\to0,\qquad
\nu_{ab}=\beta_{ab}+\ell_a(\ell_a+1)\delta_{ab},\\
V_{ab}(r)&\sim&\frac{\ell_a(\ell_a+1)}{r^2}\delta_{ab}
+\text{exp. small},\qquad\quad r\to\infty.
\end{eqnarray}

\subsection{Regular solutions, Jost solutions, Jost functions}

Apart from obvious complications related to the multi-component nature
of the coupled problem the scattering and inverse scattering theory is
very similar to the single channel case discussed in appendix \ref{appA} so
here we can be brief.

The Wronskian $W[\rho,\sigma]$ of two vectors $\rho\sim \rho_a(r)$ and
$\sigma\sim\sigma_a(r)$ is defined as
\beq
(W[\rho,\sigma])(r)=\rho_a^\prime(r)\sigma_a(r)-\rho_a(r)\sigma_a^\prime(r).
\eeq
It is constant if $\rho$ and $\sigma$ are both solutions of the
Schr\"odinger equation with the same $k^2$.

The Schr\"odinger equation (\ref{coup}) is written in the basis of partial
waves corresponding to fixed values of the orbital angular momentum $\ell_a$.
The total potential matrix $V_{ab}$ is diagonal in this basis for asymptotically
large distances. An other basis can be obtained by diagonalizing the potential
for $r\to0$:
\beq
\nu_{ab}=\sum_{A=1}^N\omega_{aA}\omega_{bA}\nu_A(\nu_A+1)
\eeq
with a real, orthogonal matrix $\omega_{aA}$ satisfying $\omega_{aA}\omega_{bA}=
\delta_{ab}$. We are interested in the case of repulsive potentials so we will
assume that the eigenvalues are positive: $\nu_A\geq0$, $A=1,...,N$.

The $N$ {\it regular solutions} $\varphi^A_k=\varphi^A(k)$ $A=1,...,N$ are
non-singular for small $r$ and have simple behaviour in this basis:
\beq
(\varphi^A_k)_a(r)\sim\omega_{aA}r^{\nu_A+1}\qquad r\to0
\eeq
and they are well-defined and analytic for all $k$ satisfying
\beq
\varphi^A(-k)=\varphi^A(k),\qquad\quad [\varphi^A(k)]^*=\varphi^A(k^*).
\eeq

The $N$ {\it Jost solutions} $f^\alpha_k=f^\alpha(k)$ $\alpha=1,...,N$
are defined by their asymptotic behaviour for large $r$:
\beq
(f^\alpha_k)_a(r)\sim\delta^\alpha_a{\rm e}^{ikr},\qquad r\to\infty
\eeq
and are well defined for ${\rm Im}\,k\geq0$ ($k\not=0$) and analytic for
$k>0$. Moreover  
\beq
[f^\alpha(k)]^*=f^\alpha(-k^*).
\eeq

The {\it Jost function (matrix)} is defined here as
\beq
f^{A\alpha}(k)=W[\varphi^A(k),f^\alpha(k)],\qquad\quad A,\alpha=1,...,N.
\eeq
It is also analytic in the upper half-plane ($k>0$), well defined for
all ${\rm Im}\,k\geq0$ ($k\not=0$) and satisfies
\beq
[f^{A\alpha}(k)]^*=f^{A\alpha}(-k^*).
\eeq
Its matrix inverse
\beq
g(k)=f^{-1}(k),\qquad\quad g^{\beta A}(k)f^{A\alpha}(k)=\delta^{\alpha\beta}
\eeq
has similar properties. The large distance asymptotics of the solution
($k\not=0$ real)
\beq
g^{\beta A}(k)\varphi^A_k=\frac{i}{2k}\left\{f^\beta_{-k}
-g^{\beta A}(k)f^{A\alpha}(-k)f^\alpha_k\right\}
\eeq
reveals that it is a superposition of an incoming free wave and an outgoing
wave:
\beq
r\to\infty:\qquad
\frac{i}{2k}\left\{\delta^\beta_a{\rm e}^{-ikr}
-\sigma^{\beta a}(k){\rm e}^{ikr}\right\}
\eeq
with scattering matrix
\beq
\sigma^{\alpha\beta}(k)=g^{\alpha A}(k)f^{A\beta}(-k),\qquad
\quad \sigma(k)=f^{-1}(k)f(-k).
\eeq
It can be shown that $\sigma(k)$ has the properties
\beq
\sigma^*(k)=\sigma(-k),\qquad
\sigma^T(k)=\sigma(k),\qquad
\sigma(k)\sigma(-k)=1,
\eeq
so it is a symmetric unitary matrix.

Similarly to the single channel Jost solution, $f^\alpha_k$ also satisfies
an integral equation which allows an analytic extension on the complex $k$
plane beyond the original domain of definition (the upper half-plane).
The rescaled Jost function
\beq
{\cal F}^\alpha_k=(-ik)^{\ell_\alpha}f^\alpha_k
\eeq
is analytic for ${\rm Im}\,k>-\kappa$ (including $k=0$). This implies that
$\sigma(k)$ is meromorphic in the strip $\vert{\rm Im}\,k\vert<\kappa$.

\subsection{The physical S-matrix}

The physical S-matrix is defined relative to the free (no potential) case:
\beq
S(k)=i^L\sigma(k)i^L,
\eeq
where $L={\rm diag}(\ell_1,...,\ell_N)$. We assume that all $\ell_a$ are
of the same parity: $(-1)^L=\pm1$. In this case the physical S-matrix $S(k)$
is also symmetric unitary for real $k$, meromorphic in the above strip
and satisfies there
\beq
S^T(k)=S(k),\qquad S^*(k)=S(-k^*),\qquad S(k)S(-k)=1.
\label{CSmat}
\eeq
In addition, if there are no zero energy bounded solutions, which we assume,
for small $k$ it can be expanded as
\beq
S^{\alpha\beta}(k)=\delta^{\alpha\beta}+{\rm O}(k^{\ell_\alpha+\ell_\beta+1}).
\label{Ck0}
\eeq

Physical phase shifts can be found by diagonalizing the S-matrix for real $k$ as
\beq
S^{\alpha\beta}(k)=\sum_{A=1}^N O_{\alpha A}(k) O_{\beta A}(k){\rm e}^{2i\delta_A(k)},
\label{Cphase}
\eeq
where $O_{\alpha A}(k)$ is a real orthogonal matrix, symmetric in $k$ and
the phase shifts $\delta_A(k)$ $A=1,...,N$ are odd functions of $k$
(modulo $\pi$).

\subsection{Bound states}

If for some $k=k_o$ the Jost function matrix is singular, ${\rm det}f(k_o)=0$,
then some linear combination of the regular solutions $\varphi^A_{k_o}$ are
also linear combination of the Jost solutions $f^\alpha_{k_o}$. This linear
combination is integrable both at $r=0$ and at $r\to\infty$ and hence a genuine
eigenvector of the Hamiltonian with eigenvalue $k_o^2$. Since this is impossible
for real $k_o$ and the Hamiltonian is hermitean with real eigenvalues, the
only possibility is that $k_o$ is imaginary. The set of solutions
\beq
k_o=i\kappa_j,\qquad j=1,...,J
\eeq
correspond to normalizable bound states. We will assume that there is a unique
(up to normalization) solution for each $\kappa_j$:
\beq
f^{(j)}_a(r)=v^\alpha_{(j)}(f^\alpha_{i\kappa_j})_a(r),\qquad\quad
\sum_\alpha(v^\alpha_{(j)})^2=1.
\eeq
This solution, and the linear combination coefficients $v^\alpha_{(j)}$ are real.
We can find the normalized bound state wave function
\beq
\psi^{(j)}_a(r)=w^\alpha_{(j)}(f^\alpha_{i\kappa_j})_a(r)
\eeq
by calculating the norming constants
\beq
\frac{1}{s_j}=\sum_{a=1}^N\int_0^\infty (f^{(j)}_a(r))^2{\rm d}r
\eeq
and putting $w^\alpha_{(j)}=\sqrt{s_j}\,v^\alpha_{(j)}$.

Like in the single channel case, we can distinguish inner and outer bound states
according to whether $\kappa_j<\kappa$ or $\kappa_j\geq\kappa$, respectively.
For inner bound states only, the corresponding pole of the scattering matrix
is related to the bound state properties:
\beq
\sigma^{\alpha\beta}(k)\sim\frac{-iw^\alpha_{(j)}w^\beta_{(j)}}{k-i\kappa_j},
\qquad
\quad ({\rm near}\ k=i\kappa_j).
\eeq

\subsection{Marchenko equation}

The derivation of Marchenko's equation for the coupled channel case proceeds
along the same line as for the single channel problem. The main difference
is that here a set of linear equations
\beq
F_{ab}(r,s)+A_{ab}(r,s)+\int_r^\infty A_{ac}(r,u)F_{cb}(u,s){\rm d}u=0,\qquad
s\geq r
\label{Cmar}
\eeq
has to be solved for the unknowns $A_{ab}(r,s)$, where
\beq
F_{cd}(u,v)=F^{\rm bound}_{cd}(u,v)+F^{\rm scatt}_{cd}(u,v)
\label{CF}
\eeq
with
\beq
F^{\rm bound}_{cd}(u,v)
=\sum_{j=1}^J w^\alpha_{(j)} w^\beta_{(j)} y^\alpha_c(i\kappa_j u)
y^\beta_d(i\kappa_j v),
\label{Fbounduv}
\eeq
\beq
\begin{split}
&F^{\rm scatt}_{cd}(u,v)=\frac{1}{2\pi i(u+v)}[\sigma^{cd}(\infty)
-\sigma^{cd}(-\infty)]\\
&+\frac{1}{2\pi}\int_0^\infty{\rm d}k\big\{[(-1)^L\delta^{\alpha\beta}-
\sigma^{\alpha\beta}(k)]y^\alpha_c(ku)y^\beta_d(kv)\\
&\qquad\quad +[(-1)^L\delta^{\alpha\beta}-
\sigma^{\alpha\beta}(-k)]y^\alpha_c(-ku)y^\beta_d(-kv)\\
&+[\sigma^{cd}(\infty)-(-1)^L\delta_{cd}]{\rm e}^{ik(u+v)}
+[\sigma^{cd}(-\infty)-(-1)^L\delta_{cd}]{\rm e}^{-ik(u+v)}\big\}.
\end{split}
\label{Fscattuv}
\eeq
In the above formulas
\beq
y^\alpha_a(x)=(i)^{\ell_\alpha+1}h_{\ell_\alpha}(x) \delta^\alpha_a.
\eeq
The infinite momentum scattering matrices are calculated to be
\beq
\sigma^{\alpha\beta}(\infty)=\sum_{A=1}^N\omega_{\alpha A}\omega_{\beta A}
{\rm e}^{-i\pi\nu_A},\qquad
\sigma^{\alpha\beta}(-\infty)=\sum_{A=1}^N\omega_{\alpha A}\omega_{\beta A}
{\rm e}^{i\pi\nu_A}.
\label{siginfty}
\eeq
These are generalizations of (\ref{inf2}).

If (and only if) all $\nu_A$ eigenvalues are integer then $\sigma(\infty)=
\sigma(-\infty)$ and this allows to simplify (\ref{Fbounduv}-\ref{Fscattuv})
as follows.
\beq
F^{\rm bound}_{cd}(u,v)=(-1)^{\frac{\ell_c+\ell_d}{2}+1}\sum_{j=1}^Jw^c_{(j)}w^d_{(j)}
h_{\ell_c}(i\kappa_j u)h_{\ell_d}(i\kappa_j v),
\label{Fbrat}
\eeq
\beq
\begin{split}
F^{\rm scatt}_{cd}(u,v)&=\frac{1}{2\pi}\int_{-\infty}^\infty{\rm d}k\big\{
[S^{cd}(k)-\delta_{cd}]h_{\ell_c}(ku)h_{\ell_d}(kv)\\
&+\sum_{A=1}^N\omega_{cA}\omega_{dA}[(-1)^{\nu_A}-(-1)^L]{\rm e}^{ik(u+v)}\big\}.
\end{split}
\label{Fsrat}
\eeq

\subsection{Levinson's theorem}

The derivation \cite{Swan} of Levinson's theorem can be generalized to
give
\beq
\int_0^\infty{\rm d}k{\rm Tr}\{S^{-1}(k)S^\prime(k)\}=2i\sum_{A=1}^N\int_0^\infty
\delta^\prime_A(k){\rm d}k,
\eeq
but this only allows us to write one \lq\lq total'' Levinson's theorem
\beq
-\frac{2}{\pi}\sum_{A=1}^N\int_0^\infty\delta^\prime_A(k){\rm d}k=
\frac{2}{\pi}\sum_{A=1}^N[\delta_A(0)-\delta_A(\infty)]=2J+\sum_{A=1}^N
\nu_A-\sum_{a=1}^N\ell_a.
\label{CLevi}
\eeq
From (\ref{siginfty}) one can conclude that there are also \lq\lq individual''
Levinson's type formulas
\beq
\delta_A(\infty)=-\frac{\pi}{2}\nu_A,\qquad\quad A=1,...,N,
\eeq
but these are valid only modulo $\pi$.

\subsection{Inverse scattering with Marchenko's method}

Inverse scattering is a method to calculate the potential matrix $q_{ab}(r)$
from the scattering data
\beq
{\cal S}=\big\{L;S^{ab}(k),-\infty<k<\infty;\{w^a_{(j)},\kappa_j\}_{j=1}^J\big\}.
\eeq
The S-matrix has to be regular and its analytic extension to the strip
$\vert{\rm Im}k\vert<\kappa$ has to be a meromorphic function satisfying the
symmetry relations (\ref{CSmat}). Its expansion around the origin
is constrained by (\ref{Ck0}), moreover after finding the eigenshifts by
the decomposition (\ref{Cphase}) the Levinson's theorem (\ref{CLevi}) must be
satisfied by nonnegative $\nu_A$ parameters. In the strip $0<{\rm Im}k<\kappa$
$S^{ab}(k)$ can have simple poles only at $k=i\kappa_j$, for all the inner
poles, with residue
\beq
-i(-1)^{\frac{\ell_a+\ell_b}{2}}w^a_{(j)}w^b_{(j)}.
\label{Cres}
\eeq

The first step of the inverse scattering procedure is the calculation
of Marchenko's $F$-functions from the scattering data by (\ref{CF},
\ref{Fbounduv},\ref{Fscattuv}). The second step is the solution of the
set of linear integral equations (\ref{Cmar}) to find $A_{ab}(u,v)$.

The third and final step is to calculate the potential using the formula
\beq
V_{ab}(r)=\frac{\ell_a(\ell_a+1)}{r^2}\delta_{ab}-2\frac{\rm d}{{\rm d}r}
A_{ab}(r,r).
\eeq
The wave functions can be obtained by the mapping
\beq
(f^\alpha_k)_a(r)=y^\alpha_a(kr)+\int_r^\infty A_{ab}(r,s)y^\alpha_b(ks){\rm d}s.
\eeq

\subsection{$N=2$ parametrization}

For applications to nucleon scattering we need the special case $N=2$.
In this case a useful parametrization is given by
\beq
S(k)=\frac{A(k)+B(k)}{2}+\frac{\big(A(k)-B(k)\big)z(k)}{4}
\rho_++\frac{A(k)-B(k)}{4z(k)}\rho_-,
\eeq 
where 
\beq
\rho_\pm=
\begin{pmatrix}
1 & \pm i\\
\pm i & -1
\end{pmatrix}.
\eeq
The symmetry of the S-matrix is already built in and the other two
requirements in (\ref{CSmat}) are satisfied if
\beq
\left.\begin{matrix}
A(k)A(-k)=1\\
A^*(k)A(k^*)=1
\end{matrix}\right\},\qquad
\left.\begin{matrix}
B(k)B(-k)=1\\
B^*(k)B(k^*)=1
\end{matrix}\right\},\qquad
\left.\begin{matrix}
z(k)=z(-k)\\
z^*(k)z(k^*)=1
\end{matrix}\right\}.
\label{Cconstraint}
\eeq
$A(k)$ and $B(k)$ are exactly like the scalar S-\lq matrix' in a single channel
problem, but the mixing part $z(k)$ is different. For real $k$,
$\delta_{1,2}(k)$ is odd in $k$ (modulo $\pi$) and $\epsilon(k)$ is even (also
modulo $\pi$). This parametrization is the
same as (\ref{para1}-\ref{para2}) after the identification
\beq
A(k)=e^{2i\delta_1(k)}, \quad B(k)=e^{2i\delta_2(k)}, \quad z(k)=e^{-2i\epsilon(k)}.
\eeq
Note that the $k\to0$ constraint (\ref{Ck0}) and the residue constraint
(\ref{Cres}) still need to be imposed.

\subsection{Rational S-matrix}

In this paper we work with rational S-matrices only so in this subsection
we apply the general theory to this special case. Rational S-matrices are
automatically meromorphic for all $k$ and satisfy
\beq
S(\infty)=S(-\infty)=S_\infty.
\eeq
If we assume (for simplicity) that all poles are simple then the S-matrix
can be written algebraically as
\beq
S^{ab}(k)=S^{ab}_\infty+\sum_n\frac{-iR^{ab}_n}{k-i\eta_n},\qquad R^{ab}_n=R^{ba}_n.
\eeq
The pole parameters $\eta_n$ are either real or come in complex conjugate
pairs: $\eta^*_n=\eta_{\bar n}$ and also the residues satisfy $(R^{ab}_n)^*=
R^{ab}_{\bar n}$.

Let the set of inner bound states be $\{\kappa_j\}$, $j=1,...,J_o$. Then
the set of pole parameters can be classified as
\beq
\{\eta_n\}=\{\kappa_j\}_{j=1}^{J_o}\cup\{\lambda_\alpha\}\cup\{\mu_\beta\},
\eeq
where ${\rm Re}(\lambda_\alpha)>\kappa_{J_o}$, ${\rm Re}(\mu_\beta)<0$ and
correspondingly
\beq
S^{ab}(k)=S^{ab}_\infty
+\sum_{j=1}^{J_o} \frac{-iR^{ab}_{(j)}}{k-i\kappa_j}
+\sum_\alpha \frac{-iR^{ab}_\alpha}{k-i\lambda_\alpha}
+\sum_\beta\frac{-i{\tilde R}^{ab}_\beta}{k-i\mu_\beta},
\eeq
where
\beq
R^{ab}_{(j)}=(-1)^{\frac{\ell_a+\ell_b}{2}}w^a_{(j)}w^b_{(j)}.
\eeq
From (\ref{Fbrat}) we get
\beq
F^{\rm bound}_{cd}(u,v)=-\sum_{j=1}^JR^{cd}_{(j)}
h_{\ell_c}(i\kappa_j u)h_{\ell_d}(i\kappa_j v)
\eeq
and from (\ref{Fsrat}), doing the integral after closing the contour in the
upper half-plane,
\beq
F^{\rm scatt}_{cd}(u,v)=
\sum_{j=1}^{J_o}R^{cd}_{(j)}
h_{\ell_c}(i\kappa_j u)h_{\ell_d}(i\kappa_j v)
+\sum_\alpha R^{cd}_\alpha
h_{\ell_c}(i\lambda_\alpha u)h_{\ell_d}(i\lambda_\alpha v).
\eeq
We see that the contribution of the inner poles completely cancels. The
final result for Marchenko's $F$-function takes the form
\beq
F_{cd}(u,v)=\sum_m R^{cd}_m h_{\ell_c}(i\xi_m u)h_{\ell_d}(i\xi_m v),
\label{Fxi}
\eeq
where the set of relevant poles is
\beq
\{\xi_m\}=
\{\kappa_j\}_{j=J_o+1}^J\cup\{\lambda_\alpha\}.
\eeq

\subsection{Algebraic solution}

The advantage of using Marchenko's method for rational S-matrices is that
the solution can be obtained algebraically \cite{vonGeramb1,vonGeramb2}.
Introducing the notation
\beq
\omega^{(a)}_n(r)=h_{\ell_a}(i\xi_n r)
\eeq
(\ref{Fxi}) reads
\beq
F_{cd}(u,v)=\sum_n R^{cd}_n\omega^{(c)}_n(u)\omega^{(d)}_n(v).
\eeq
From the Marchenko equation (\ref{Cmar}) we first obtain that the solution
must be of the form
\beq
A_{ab}(r,s)=-\sum_n {\cal V}^{ab}_n(r)\omega^{(b)}_n(s)
\eeq
and then the equation can be reduced to a set of linear algebraic equations
for the unknowns ${\cal V}^{ab}_n(r)$:
\beq
\sum_{m,c}{\cal V}^{ac}_m(r){\cal M}^{cb}_{mn}(r)=\omega^{(a)}_n(r)R^{ab}_n.
\eeq
The matrix of the linear problem is
\beq
{\cal M}^{cb}_{mn}(r)=\delta^{cb}\delta_{mn}+I^{(c)}_{mn}(r)R^{cb}_n,
\eeq
where
\beq
I^{(c)}_{mn}(r)=\int_r^\infty{\rm d}u\,\omega^{(c)}_m(u)\omega^{(c)}_n(u).
\eeq
This integral can be explicitly evaluated using (\ref{Hint}-\ref{Hint2}).

Introducing the matrix inverse ${\cal N}^{bd}_{nk}(r)$ satisfying
\beq
\sum_{n,b}{\cal M}^{cb}_{mn}(r){\cal N}^{bd}_{nk}(r)=\delta^{cd}\delta_{mk}
\eeq
the solution for ${\cal V}^{ad}_k(r)$ is
\beq
{\cal V}^{ad}_k(r)=\sum_{n,b}\omega^{(a)}_n(r)R^{ab}_n{\cal N}^{bd}_{nk}(r)
\eeq
and finally we obtain
\beq
A_{ab}(r,s)=-\sum_{n,k,c}\omega^{(a)}_n(r)R^{ac}_n{\cal N}^{cb}_{nk}(r)
\omega^{(b)}_k(s).
\eeq


  


\vfill\eject

\end{document}